\documentclass[12pt]{article}

\usepackage{amsthm,latexsym,multibox,amssymb,wasysym,textcomp,stmaryrd}
\input amssym.def
\input amssym.tex

\def\g{\gamma}
\def\d{\delta}

\def\l{\lambda}

\def\m{\mu}
\def\n{\nu}
\def\r{\rho}

\def\s{\sigma}

\def\z{\zeta}

\def\be{\begin{equation}}
\def\ee{\end{equation}}
\def\bqn{\begin{eqnarray}}
\def\eqn{\end{eqnarray}}
\def\nn{\nonumber}

\def\cu{{\cal U}}

\newtheorem{theorem}{Theorem}
\newtheorem{lemma}{Lemma}
\newtheorem{corollary}{Corollary}

\begin{document}

\begin{flushright}
IHES/P/05/15 \\
{\tt hep-th/0505068}
\end{flushright}
\vskip 2cm

 \begin{centering}

{\large {\bfseries Gauge invariants and Killing tensors\\\vspace{1mm} in
higher-spin gauge theories}}

 \vspace{2cm}
Xavier Bekaert$^\clubsuit$\footnote{E-mail address: \tt
bekaert@ihes.fr}
and Nicolas Boulanger$^{\eighthnote}$\footnote{Charg\'e de Recherches FNRS (Belgium), E-mail address: \tt nicolas.boulanger@umh.ac.be}\\
 \vspace{.8cm}
 {\small $^\clubsuit$
 Institut des Hautes \'Etudes Scientifiques, Le Bois-Marie\\
 35 route de Chartres, 91440 Bures-sur-Yvette (France)\\
  \vspace{.3cm}$^{\eighthnote}$ Universit\'e de Mons-Hainaut, M\'ecanique et Gravitation
\\ 6 Avenue du Champ de Mars, 7000 Mons (Belgium)
  }     \\

\vspace{1.2cm}

\end{centering}

\begin{abstract}
In free completely symmetric tensor
gauge field theories on Minkowski space-time, all gauge invariant functions and
Killing tensor fields are computed, both on-shell and
off-shell. These problems are addressed in the metric-like formalisms.
\end{abstract}

\vspace{1.5cm}


\vspace{.5cm}


\vspace{4.5cm}

\pagebreak

\section{Introduction}

The old Fronsdal's programme -- that consists in introducing
consistent interactions among massless higher-spin fields
\cite{Fronsdal:1978rb} -- still remains an important open
mathematical problem of classical field theory. Numerous
preliminary results toward this goal have been obtained (see e.g.
the review paper \cite{BCIV} and references therein) which reveal
surprising properties of higher-spin gauge fields. A better
(``geometrical") understanding of the gauge structure
underlying these theories will presumably be necessary for
actually completing Fronsdal's programme. As a very preliminary
step in this direction, some properties of the Abelian gauge
transformations of the free theories are investigated closely
in the present paper. More precisely, we focus on the following
problems: the determination of all gauge invariant functions and
all Killing tensor fields of free completely symmetric tensor
gauge field theories in flat background, respectively on-shell and
off-shell, under the sole assumption of locality.
These problems are solved in both constrained and unconstrained metric-like approaches.

The plan of the paper is as follows. Free higher-spin gauge
theories are reviewed in Section \ref{statement} where the
problems addressed are formulated in a more precise way. As explained in Section \ref{reformulation}, the
problems are solved via a cohomological reformulation in the Becchi-Rouet-Stora-Tyutin (BRST)
formalism and their general solutions are contained in the corollary of Theorem
\ref{Hgamma} and in Theorem \ref{Killing} which are given in the
section \ref{results} that concludes the paper. In order to spell out the notation and
make the paper as self-contained as possible, Appendix
\ref{irreps} reviews some textbook material on
irreducible representations of the general linear and orthogonal Lie algebras. Theorems
\ref{Hgamma} and \ref{Killing} are respectively proved in
Appendices \ref{proof1} and \ref{proof2}.

\section{Statement of the problems}\label{statement}

Einstein's gravity theory is a non-Abelian massless spin-$2$ field
theory, the two main formulations of which are the ``metric" and
the ``frame" approaches. In a very close analogy, there exist two
main approaches to higher-spin ({\it i.e.} spin $s>2$) field
theories that are by-now referred to as ``metric-like"
\cite{Fronsdal:1978rb,deWit:1979pe} and ``frame-like"
\cite{BCIV,Vframe}. The present paper essentially deals with the
former approach where the massless field is represented by a completely symmetric
field $\varphi$ of rank\footnote{Throughout this text,
the spin $s$ is taken to be a strictly positive
integer. Note that the half-integer spin case is also covered
here since all the results apply to fermions by simply
replacing the spin $s$ by its integer part $[s]$.} $s>0$, the gauge
transformation of which reads
\begin{equation}
\delta_\varepsilon\varphi_{\m_1\ldots\,\,\m_s}=s\,\partial_{(\m_1}\varepsilon_{\m_2\ldots\m_s)}\,,\label{gtransfo}
\end{equation}
where the curved bracket denotes complete symmetrization with
strength one\footnote{For example,
$\varphi_{(\m_1\ldots\,\,\m_s)}\equiv
\varphi_{\m_1\ldots\,\,\m_s}$.} and the Greek indices run over $n$
values ($n\geq 3$). The gauge parameter $\varepsilon$ is a
completely symmetric tensor field of rank $s-1$. For spin $s=2$,
the gauge field $\varphi_{\mu\nu}$ represents the graviton while
the gauge transformations (\ref{gtransfo}) correspond to
linearized diffeomorphisms.

Originally, some algebraic constraints were imposed on the
metric-like gauge field  and gauge parameters
\cite{Fronsdal:1978rb,deWit:1979pe}. More precisely, given
$\eta_{\m\n}$ the metric tensor of the flat
background\footnote{Our results are independent of the choice of
signature because they rely on algebraic considerations
only. For physical reasons, we choose the Lorentzian signature.}, the gauge parameter $\varepsilon$ was taken to be
``traceless" ($s\geq 3$) and the gauge field $\varphi$ to be
``double-traceless" ($s\geq 4$).

{\bfseries Notation:} The trace of any tensor $T_{\m_1\ldots
\m_r}$ is written as
$T^\prime_{\mu_1...\mu_{r-2}}$($\equiv\eta^{\n_1\n_2}T_{\mu_1...\mu_{r-2}\n_1\n_2}$).
Thus the previous ``trace" constraints are written as
\begin{equation}\varepsilon^{\prime}_{\mu_1...\mu_{s-3}}=0\,,\quad\varphi^{\prime\prime}_{\mu_1...\mu_{s-4}}=0\,.\label{traceconstr}\end{equation}
The traceless part of any tensor $T_{\m_1\ldots \m_r}$ is denoted
by $\widehat{T}_{\m_1\ldots \m_r}$ (that is
${\widehat{T}}_{\m_1\ldots \m_{r-2}}^{\prime}=0$). Analogously,
the ``double traceless" part of $T$ is denoted by
$\widetilde{T}_{\m_1\ldots \m_r}$ ($\widetilde{T}^{\prime\prime}_{\m_1\ldots \m_{r-4}}=0$). Hence,
the constraints (\ref{traceconstr}) can also be written
$$\varepsilon_{\mu_1...\mu_{s-1}}=\widehat{\varepsilon}_{\mu_1...\mu_{s-1}}\,,\quad\varphi_{\mu_1...\mu_s}=\widetilde{\varphi}_{\mu_1...\mu_s}\,.$$

Recently \cite{Francia:2002aa}, Francia and Sagnotti realized that
the algebraic constraints (\ref{traceconstr}) could be
consistently relaxed in the sense that the former theory appears
as a gauge fixing of a consistent non-local theory without any
trace constraint. The former and latter formulations will be
referred to as ``constrained" and ``unconstrained" approaches (see
\cite{Bouatta:2004kk} for pedagogical reviews on both of them).
Though the absence of the constraints (\ref{traceconstr})
simplifies our algebraic analysis, the non-locality property of
the unconstrained approach is still rather elusive, therefore both
cases will be treated for the sake of completeness.

\subsection{Gauge invariant functions}

There is a general belief that a local function of the completely
symmetric gauge field $\varphi_{\m_1\ldots\,\,\m_s}$ is gauge
invariant under unconstrained gauge transformations if and only if
it depends on $\varphi$ only via the de Wit--Freedman curvature
tensor field \cite{deWit:1979pe}\begin{equation}
{\cal
R}_{\m_1\ldots\,\,\m_s\,,\,\n_1\ldots\,\,\n_s}
:=\partial_{\m_1}\ldots\partial_{\m_s}\varphi_{\n_1\ldots\,\,\n_s}+\ldots\end{equation}
and its partial derivatives. The dots stand for the terms
necessary for the tensor $\cal R$  to belong to the irreducible
${\mathfrak{gl}}(n)$-module associated with the rectangular two-row
Young diagram of length $s$ depicted
\begin{picture}(60,15)(-5,2)
\multiframe(0,7.5)(13.5,0){1}(50,7){}\put(53,7){$s$}
\multiframe(0,0)(13.5,0){1}(50,7){}\put(53,0){$s$}\end{picture}\hspace{.3cm},
so that $\cal R$ satisfies the algebraic identities$${\cal
R}_{\m_1\ldots\,\,\m_s\,,\,\n_1\ldots\,\,\n_s}={\cal
R}_{(\m_1\ldots\,\,\m_s)\,,\,\n_1\ldots\,\,\n_s}={\cal
R}_{\m_1\ldots\,\,\m_s\,,\,(\n_1\ldots\,\,\n_s)}\,,$$ $${\cal
R}_{(\m_1\ldots\,\,\m_s\,,\,\n_1)\n_2\ldots\,\,\n_s}=0\,.$$ For
constrained fields and parameters, the most general local gauge invariant
function of $\varphi$ is believed to depend only on the de
Wit--Freedman curvature tensor $\cal R$ and on the Fronsdal tensor
\begin{equation}\label{Fronsdaltensor}{\cal
F}_{\m_1\ldots\,\,\m_s}\,:=\,\Box\,\varphi_{\mu_1...\mu_s}\,-\,s\,\partial^\n\partial_{(\m_1}\varphi_{\mu_2...\mu_s)\,\n}
\,+\,{s(s-1)\over 2
}\,\partial_{(\m_1}\partial_{\m_2}\varphi^\prime_{\mu_3...\mu_s)\n_1\n_2}
\end{equation}
together with their derivatives.

We would like to put the previous expectations on firm
mathematical grounds. Part of the present paper (Section
\ref{Th1}) is therefore mainly devoted to a mathematically
rigorous analysis of the following problem: the determination of
all possible gauge invariant local functions depending on the
gauge field $\varphi$ and a finite number of its partial
derivatives both in the unconstrained and constrained approaches.
Actually, the answer depends on whether the system is studied
on-shell or not since, for instance, \begin{equation}{\cal
F}_{\m_1\ldots\,\,\m_s}\,\approx
0\label{Fronsdalequ}\end{equation} in the constrained approach.
The symbol $\approx$ means that the equality holds on-shell. By
combining the Damour-Deser identity \cite{DD} (which relates the
trace of $\cal R$ to $s-2$ exterior derivatives of $\cal F$) with
the generalized Poincar\'e lemma \cite{DuboisV}, we argued in
\cite{BB2} that the field equation
\begin{equation}{\cal
R}^\prime_{\m_1\ldots\,\,\m_s\,,\,\,\n_1\ldots\,\,\n_{s-2}}\,\approx
0\label{Weylequ}\end{equation} in the unconstrained approach is
dynamically equivalent to (\ref{Fronsdalequ}) in the constrained
approach. More precisely,
\begin{eqnarray}{\cal
F}_{\m_1\ldots\,\,\m_s}(\widetilde{\varphi})&=& 0\,,\quad
\delta\widetilde{\varphi}_{\m_1\ldots\,\,\m_s}=s\,\partial_{(\m_1}\widehat{\varepsilon}_{\m_2\ldots\m_s)}\nn\\
&&\Longleftrightarrow \quad{\cal
R}^\prime_{\m_1\ldots\,\,\m_s\,,\,\,\n_1\ldots\,\,\n_{s-2}}(\varphi)=
0\,,\quad
\delta\varphi_{\m_1\ldots\,\,\m_s}=s\,\partial_{(\m_1}\varepsilon_{\m_2\ldots\m_s)}\,.\label{dynequiv}
\end{eqnarray} The spin-$s$ Weyl-like tensor $\cal W$ is defined as the
traceless part of the spin-$s$ tensor $\cal R$, $${\cal
W}_{\m_1\ldots\,\,\m_s\,,\,\n_1\ldots\,\,\n_s}:=\widehat{{\cal
R}}_{\m_1\ldots\,\,\m_s\,,\,\n_1\ldots\,\,\n_s}\,,$$ hence it
belongs to the irreducible $\mathfrak{o}(n-1,1)$-module associated
with the rectangular two-row Young diagram of length $s$. In both
approaches, the on-shell de Wit -- Freedman tensor is equal to the
Weyl-like tensor
$${\cal
R}_{\m_1\ldots\,\,\m_s\,,\,\n_1\ldots\,\,\n_s}\approx{{\cal
W}}_{\m_1\ldots\,\,\m_s\,,\,\n_1\ldots\,\,\n_s}\,.$$ A
prerequisite of Vasiliev's unfolded approach is the determination
of all on-shell nontrivial derivatives of the fieldstrengths.
The spin-$s$ de Wit -- Freedman tensor obeys some differential
Bianchi identity and, on-shell, it is traceless, divergenceless
and satisfies the massless Klein-Gordon equation \cite{BB2}. This
implies that the $k$th partial derivatives of the spin-$s$ de Wit
-- Freedman tensor belong off(on)-shell to the irreducible
$\mathfrak{gl}(n)$($\mathfrak{o}(n-1,1)$)-module labeled by the
two-row Young diagram
\begin{picture}(115,15)(0,2)
\multiframe(10,7.5)(13.5,0){1}(65,7){}\put(80,7.5){$s+k$}
\multiframe(10,0)(13.5,0){1}(35,7){}\put(50,0){$s$}
\end{picture}. They span
the so-called ``twisted adjoint module" and play a fundamental
role in the unfolded approach.\vspace{1mm}

To address the previously-stated problem of determining the
gauge-invariant local functions of the (un)constrained symmetric
gauge field $\varphi$, we reformulate it as a cohomological
problem in the BRST formalism (see {\it e.g.} \cite{Henneauxbook}
for a comprehensive review). More precisely, we use the fact that
the set of such gauge invariant functions is given by the local BRST
cohomology group $H^0(s)$ in vanishing ghost number in the sector
without antifields. This cohomology group is actually identical to
the cohomology group $H^0(\g)$ of the longitudinal exterior
differential $\g$ in the sector without antifields. In the present
paper, the complete local cohomology of $\g$ is determined. This result
is useful because the knowledge of $H^*(\g)$ is the
first ingredient in the computation of the local BRST cohomology
group $H^{n,0}(s|\,d)$ in top form degree and in vanishing ghost
number for higher-spin gauge theories, a subject which will be
addressed later in the context of the classification of local
consistent vertices for higher-spin gauge theories in flat space-time
\cite{BBetal}.

These results will be compared to two other cohomological analyses
of somewhat related problems: the generalized Poincar\'e lemma of
\cite{DuboisV} (extended to arbitrary irreducible tensors under
the general linear group in \cite{BB1}) and the problem of writing
down non-trivial equations in the unfolded formalism \cite{Shaynk}
(see also Sections 9 and 10 of \cite{BCIV} for a review).

\subsection{Killing tensors}\label{Killingsect}

Another problem of physical interest for a better understanding of
the higher-spin symmetries is the determination of all Killing
tensor fields on Minkowski space-time, that is, the symmetric
tensor fields satisfying the following (off-shell) Killing-like
equation \cite{Eisen}
\begin{equation}\partial_{(\m_1}\varepsilon_{\m_2\ldots\m_s)}=0\,.\label{Killingequ}\end{equation}
Actually this problem is easy to solve for tensor fields
$\varepsilon_{\m_1\ldots\m_{s-1}}(x)$ which are formal power
series in $x$. Killing tensor fields on spaces of constant
curvature have been extensively studied by mathematicians (see
{\it e.g.} \cite{Killingconstcurvv,Killingtensorsolution}).

A more involved problem is the determination of all local
``on-shell Killing tensor fields". Non-trivial on-shell Killing
tensor fields on flat space-time are solutions of the conditions
\begin{equation}
\partial_{(\m_1}\varepsilon_{\m_2\ldots\m_s)}\approx
0\,,\quad\quad \varepsilon_{\m_1\ldots\m_{s-1}}\not\approx 0\,,
\label{weaK} \end{equation} which define a cohomological problem.
Again, in order to address this problem we formulate it in the
context of the local BRST formalism, by using the fact that these
local non-trivial on-shell Killing tensor fields are in one-to-one
correspondence with cocycles of the local Koszul-Tate cohomology
group $H^n_2(\delta|d)$ in top form degree and antifield number
two \cite{Barnich:2000zw}. Though interesting in its own sake, the
knowledge of $H^n_2(\delta|d)$ is the second ingredient in the
computation of the local BRST cohomology group $H^{n,0}(s|d)$
\cite{BBetal}.

Generally speaking, the global symmetries of a solution of some
field equation correspond to the space of gauge parameters leaving
the gauge fields invariant under gauge transformations evaluated
at the solution. Furthermore, for the flat vacuum solution they
are expected to correspond to the full symmetry algebra of the
theory. More specifically, the Killing tensors (\ref{Killingequ})
of the infinite tower of higher-spin fields should be related to a
higher-spin algebra (if any) in flat space-time. Indeed, we prove
in Corollary \ref{coro3} (Subsection \ref{KTate}) that they are in
one-to-one correspondence with the elements of a Minkowski
higher-spin algebra that can be obtained as a quotient of the
universal enveloping algebra $\cu(\mathfrak{iso}(n-1,1))$ of the
Poincar\'e algebra or as an In\"{o}n\"{u}-Wigner contraction of
the anti de Sitter / conformal higher-spin algebras of Eastwood
and Vasiliev \cite{Eastwood,Valgebra} in the flat limit
$\Lambda\rightarrow 0$.

These higher-spin gauge symmetry algebras might eventually find
their origin in the general procedure of ``gauging" some global
higher-symmetry algebras of free theories, as we argue now.
Reformulating an observation of \cite{Berends}, the action of a
complex bosonic field $\phi_{\m_1\ldots\m_t}$ with arbitrary spin
$t$ is invariant under the global infinitesimal transformations
\begin{equation}\d_\lambda\phi_{\m_1\ldots\m_t}\,=\,
i^{s-2}\,\lambda_{\n_1\ldots\n_{s-1}}\partial^{\n_1}\ldots\partial^{\n_{s-1}}
\phi_{\m_1\ldots\m_t}\,,\label{univtranslationalg}\end{equation}
where $\lambda_{\n_1\ldots\n_{s-1}}$ are completely symmetric
\emph{constant real} tensors. For $s=1$, one recovers the usual
infinitesimal $\mathfrak{u}(1)$ phase transformation and, for
$s=2$, one obtains the usual infinitesimal action of the
translation group ${\mathbb R}^d$. If one tried to gauge the
global symmetry transformations of the form
(\ref{univtranslationalg}) by replacing the constant parameters
$\lambda_{\n_1\ldots\n_{s-1}}$ by arbitrary completely symmetric
tensor fields $\varepsilon_{\n_1\ldots\n_{s-1}}$, then the usual
prescription would require the introduction of some connection in
order to define proper covariant derivative. It is suggestive to
interpret the gauge fields $\varphi_{\m_1\ldots\m_s}$ as entering
into the definition of the connection so that the Abelian gauge
transformations (\ref{gtransfo}) should somehow correspond to the
linearization of the non-Abelian gauge transformations. Moreover,
the trace condition on the gauge parameter may find a natural
explanation according to this line of thinking. For a scalar field
($t=0$), the infinitesmal transformation
(\ref{univtranslationalg}) reads
\begin{equation}\d_\lambda\phi\,\approx\,
i^{s-2}\,\widehat{\lambda}_{\n_1\ldots\n_{s-1}}\partial^{\n_1}\ldots\partial^{\n_{s-1}}\phi\,,
\label{univtranslationalgonsh}\end{equation} that is, the
parameters may be assumed to be traceless without loss of
generality, since $(\Box-m^2)\phi\approx 0$. This is in agreement
with the fact that trace conditions are present in the so-called
``on-shell" higher-spin algebra and not in the ``off-shell" one
(see Section 5 of \cite{BCIV}).\vspace{1mm}

Let $\{P^\m\}$ be a basis of the (Abelian) translation algebra
${\mathbb R}^n$ that is represented by the Hermitian operators
$\textsc{P}^\m=i\partial^\m$ acting on $\phi_{\m_1\ldots\m_t}$.
The Hermitian kinetic operator $\textsc{K}(\textsc{P})$ for the
field $\phi_{\m_1\ldots\m_t}$ is an element of the polynomial
algebra ${\mathbb R}[\textsc{P}^\m]$. The translation operators
$-i\lambda_\m\textsc{P}^\m$ obviously generate symmetries of the
quadratic action $\int \langle\phi |\textsc{K}|\phi\rangle$ since
they are anti-Hermitian and commute with the kinetic operator, but the
same is true for any product of such translation operators,
therefore the universal enveloping algebra of the translations
generates an infinite-dimensional algebra of global symmetries of
the free theory. The universal enveloping algebra of an Abelian
Lie algebra is the corresponding polynomial algebra, in our case
${\cal U}({\mathbb R}^n)\cong{\mathbb R}[P^\m]$. This algebra is
unitarily represented by the anti-Hermitian operators
$-i\lambda_{\n_1\ldots\n_{s-1}}\textsc{P}^{\n_1}\ldots
\textsc{P}^{\n_{s-1}}$ acting on the field $\phi_{\m_1\ldots\m_t}$
as in (\ref{univtranslationalg}). (The symmetry algebra ${\mathbb
R}[\textsc{P}^\m]$ should be quotiented by the ideal of elements
proportional to $\textsc{K}(\textsc{P})$ in order to obtain the algebra of
non-trivial symmetries of the action. For $t=0$ and $\textsc{K}=\textsc{P}^2+m^2$,
the action of the latter algebra on the scalar $\phi$ is written
explicitly in (\ref{univtranslationalgonsh}).)
The Abelian Lie algebra\footnote{$\tt hs$ stands at the same time
for ``higher-spin" and ``higher-symmetry".}
${\tt hs}({\mathbb R}^n)$ of the higher-symmetries
(\ref{univtranslationalg}) is obtained from the associative
algebra ${\cal U}({\mathbb R}^n)$ with the Lie
bracket given by the commutator. The symmetry group ${\tt
HS}({\mathbb R}^n)$ is obtained by exponentiation of ${\tt
hs}({\mathbb R}^n)$ with generic element $U$ in the unitary representation: $$U=\exp(-i\lambda_{\n_1\ldots\n_{s-1}}\textsc{P}^{\n_1}\ldots
\textsc{P}^{\n_{s-1}})\,.$$ This leads to the finite form
\begin{equation}\langle x|\phi\rangle\,\longrightarrow\, \langle x|\,U|\phi\rangle=\int d^n y\,
U(x-y)\,\langle y|\phi\rangle=\exp(-i\l)\langle
x|\phi\rangle+\langle
x+\l\,|\phi\rangle+\ldots\label{finitetransfo}\end{equation} of
the infinitesimal transformations (\ref{univtranslationalg}) where
$$U(z)\,=\,(2\pi)^{-n}\int
 d^np\,\,\exp\, [-i\,(\,z^\mu
p_\mu+\,\lambda_{\n_1\ldots\n_{s-1}}\,p^{\n_1}\ldots
p^{\n_{s-1}})]$$ and the dots in (\ref{finitetransfo}) denote the transformations corresponding to $s\geq 3$.
These new terms are readily seen to be non-local (in the sense that they cannot be written in the form $\langle x^\prime|\phi\rangle$ where $x^\prime=f(x)$) because $U(z)$ is a delta of Dirac only when $s=1,2$.

The previous reasoning entirely applies to the more complicated
Poincar\'e algebra $\mathfrak{iso}(n-1,1)$ case and leads to
some Minkowski higher-spin algebra.
One should point out that the previous arguments provide a mean
to evade the conclusions of the S-matrix no-go theorems against
higher space-time symmetries \cite{cm}, even in flat space-time.
(For a more general discussion of higher-symmetries along the
lines of this subsection, see \cite{B}.)

\section{Cohomological reformulation of the
problems}\label{reformulation}

\subsection{Jet space}

To reformulate a field theoretical problem -- a functional problem --
into a finite-dimensional algebraic problem -- much more easy to
cope with -- one usually treats the fields and their partial
derivatives as independent coordinates of a so-called ``jet space".
(We follow closely the terminology of the report \cite{Barnich:2000zw}.)

{\bfseries Notation:} Frequently, we will omit indices for fields
whose rank has been defined previously. The letter $\Phi$
will collectively denote the set of field variables (some of which
may be Grassmannian). The $p$th partial derivatives of the
variables $\Phi$ will be denoted by $\partial^p \Phi$, that is
$\partial^p\Phi\,\sim\,\partial_{\m_1}\ldots\partial_{\m_p}\Phi\,.$\vspace{1mm}

An (off-shell) {\textit{local function of the field variables}}
$\Phi$ is a function $f=f(x,[\Phi])$ of the space-time
coordinates, polynomial in the field variables $\Phi$ and a finite number of
their derivatives. The notation $[\Phi]$ means dependence on the
variables $\Phi$, $\partial\Phi$, $\partial^2\Phi$, ...,
$\partial^k\Phi$ for some finite but otherwise arbitrary integer $k$. The
{\textit{jet space of order}} $k$ is defined as the direct product
$J^k(E)={\cal M}\times V^k$, where $\cal M$ is the $n$-dimensional
space-time manifold, $V^k$ the space with coordinates given by
$\Phi$, $\partial\Phi$, $\partial^2\Phi$, ..., $\partial^k\Phi$
and $E\equiv J^0(E)$ the jet space of order zero with coordinates
$x$ and $\Phi$. A local function is thus a function on a jet space
of some finite order (which will always be omitted in the sequel
for the sake of readability), that is, an element of the space
of sections on the trivial bundle $J^*(E)$. We
will denote the supercommutative algebra of (off-shell) local
functions of the field variables $\Phi$ by $\Upsilon_0(\Phi)$.

Let $\Psi\in\Upsilon_0(\Phi)$ be the left-hand-side of the
equations of motion $\Psi(\,[\Phi]\,)\approx 0$. The collection of
equations $[\Psi]= 0$ defines a submanifold of the jet space
$J^*(E)$ called the {\textit{stationary surface}} and denoted
by $\Sigma$. It turns out to be convenient to impose the following
{\emph{regularity conditions}}: (i) the local functions $[\Psi]$
can be split into independent and dependent ones and (ii) the
independent functions can be locally taken as the first
coordinates of a new, regular, coordinate system on the jet space
$J^*(E)$ in the vicinity of the stationary surface $\Sigma$. The
algebra of local functions proportional to the variables $[\Psi]$ forms an
ideal, which we denote by ${\cal I}$, that leads to the
equivalence relation
$$f,g\in \Upsilon_0(\Phi)\,;\quad\quad f\approx g\quad\Longleftrightarrow \quad f-g\in{\cal I}\,.$$
One defines the algebra $\Upsilon_\Psi(\Phi)$ of {\textit{on-shell
local functions}} as the quotient of the algebra
$\Upsilon_0(\Phi)$ by the ideal ${\cal I}$. The regularity
conditions imply that the space of on-shell local functions is
isomorphic to the space $\Gamma(\Sigma)$ of local functions on the
stationary surface $\Sigma$.

We will not consider topologically non-trivial space-time and/or field manifolds.
Therefore a {\textit{field history}} is
defined as a map $h:{\cal{M}}\rightarrow E:x\mapsto (x,\Phi(x))$.
(In the more general case, one should now introduce sections
instead of functions, as well as jet bundles.) Any field history
naturally induces a map from $\cal M$ to any given jet space
$J^k(E)$ by evaluating the partial derivatives at each point of
$\cal M$. The evaluation of a local function at a field history
yields a {\textit{local space-time function}}.

\subsection{Fields and antifields}

It is straightforward to apply the antifield BRST
formalism\footnote{For reviews on the Lagrangian ({\it i.e.}
antifield) BRST method, we refer the reader to the book
\cite{Henneauxbook} and the lectures \cite{Henneaux:1989}.} to
free ``irreducible" gauge theories, such as Fronsdal's theory describing
rank-$s$ completely symmetric gauge fields. Therefore we
directly give the main results.

Let us start with the unconstrained gauge theory. The gauge
transformations (\ref{gtransfo}) are irreducible, hence it is
sufficient to introduce a fermionic ghost field
$\xi_{\m_1\ldots\m_{s-1}}$ for each bosonic gauge parameter
$\varepsilon_{\m_1\ldots\m_{s-1}}$. The ``field'' content is
thereby enlarged to
$$
\Phi=\{\,\varphi\,,\,\xi\}\,,\quad\quad\epsilon(\varphi)=0\,,\quad\epsilon(\xi)=1\,,
$$
where $\epsilon(Z)$ denotes the Grassmann parity of the field
$Z$. The corresponding algebra $\Upsilon_0(\Phi)$ of local
functions of the fields is $\mathbb N$-graded by the pureghost
number. The corresponding diagonal operator $puregh$ is an even
derivation defined by the following grading of the generators
$$puregh(\varphi)=0,\quad puregh(\xi)=1.$$
To each field $Z\in\Phi$ we associate an antifield $Z^*$ of
opposite parity. The set of associated antifields is then
$$
\Phi^*=\{\,\varphi^{*}\,,\,\xi^*\}\,,\quad\quad\epsilon(\varphi^*)=1\,,\quad\epsilon(\xi^*)=0\,.
$$
The pure ghost number of any antifield vanishes. The algebra
$\Upsilon(\Phi,\Phi^*)$ of local functions of the fields and
antifields is bigraded: first by the pureghost number, second by
the antighost number defined from
$$antigh(\varphi)=0,\quad antigh(\xi)=0\quad antigh(\varphi^*)=1,\quad antigh(\xi^*)=2.$$
The (total) ghost number equals the difference between the pure
ghost number and the antighost number$$gh(Z)=puregh(Z)-antigh(Z)\,,\quad\forall Z\in \Phi\cup\Phi^*\,.$$

The previous paragraph similarly applies to the constrained
gauge theory, the only difference being that some trace
constraints must be imposed. The field content of the
constrained theory is
$$
\overline{\Phi}=\{\,\widetilde{\varphi}\,,\,\widehat{\xi}\,\}\,,
\quad\quad\epsilon(\widetilde{\varphi})=0\,,\quad\epsilon(\widehat{\xi}\,)=1\,.
$$
The set of associated antifields is then
$$
\overline{\Phi}^*=\{\,\widetilde{\varphi}^{*}\,,\,\widehat{\xi}^*\}\,,\quad\quad\epsilon(\widetilde{\varphi}^*)=1\,,\quad\epsilon(\widehat{\xi}^*)=0\,.
$$

Following the comments made in the introduction, the stationary
surface is the submanifold defined by ${{\cal
R}^\prime}(\,[\varphi]\,)\approx 0$ in the unconstrained approach and by
${{\cal F}}(\,[\widetilde{\varphi}]\,)\approx 0$ in the constrained
approach. Thus the algebras of on-shell local functions are
respectively $\Upsilon_{{\cal R}^\prime}(\varphi)$ in the
unconstrained approach and $\Upsilon_{{\cal
F}}(\widetilde{\varphi})$ in the constrained approach.

\subsection{BRST differential}

The {\textit{BRST differential}} $s$ acts on the algebra
$\Upsilon_0(\Phi,\Phi^*)$ and its ghost number is equal to one.
Its action obviously induces an action on the subalgebra
$\Upsilon_0(\overline{\Phi},\overline{\Phi}^*)$. It decomposes
according to the antighost number as
$$s=\d + \gamma \,,\quad antigh(\d)=-1\,,\quad antigh(\g)=0$$
and provides the algebra of gauge invariant functions on the stationary surface through its cohomology at
ghost number zero $H^0(s)$. The {\textit{Koszul-Tate
differential}} $\delta$ acts trivially on the fields, $\d\Phi=0$.
In the antifield sector, $\d\varphi^*$ is equal to the equations
of motion and $\d \xi^*$ is proportional to the ``Noether
identities". The {\textit{exterior derivative along the gauge
orbits}} is the differential $\gamma$ defined by replacing the
gauge parameters by the corresponding ghosts in the gauge
transformation (\ref{gtransfo})
\begin{equation}\label{gammaformula}
\g\varphi_{\m_1\ldots\,\,\m_s}=s\,\partial_{(\m_1}\xi_{\m_2\ldots\m_s)}\,,\quad\g
\xi_{\m_1\ldots\m_{s-1}}=0\,,
\end{equation}

On the one hand, the Koszul-Tate differential $\d$ implements the
restriction on the stationary surface $\Sigma$ (since it provides
a resolution of the algebra of functions on the stationary
surface). On the other hand, the longitudinal exterior
differential $\g$ picks out the gauge-invariant functions via its
cohomology at pureghost number zero.
The cohomological groups $H^0(s,\Upsilon_0(\Phi,\Phi^*))\cong H(\g,\Upsilon_{{\cal R}^\prime}(\varphi))$ and $H^0(s,\Upsilon_0(\overline{\Phi},\overline{\Phi}^*))\cong H(\g,\Upsilon_{{\cal
F}}(\widetilde{\varphi}))$ can be read off Corollary \ref{corol} given in the next
section.

\section{The results and their physical interpretations}\label{results}

\subsection{Longitudinal exterior cohomology}\label{Th1}

Let us define the Grassmann algebra $\Xi$ generated by the
fermionic variables $d^{k}\xi$, where $k$ is a positive integer
not larger than $s-1$ and where each exterior derivative $d$ in
the power $d^{k}$ acts on a different space-time index of $\xi$.
It is easy to see that the tensors $d^{k}\xi$ belong to the
irreducible $\mathfrak{gl}(n)$-module labeled by the Young diagram
\begin{picture}(80,15)(-5,2)\put(58,6.5){s\,--{\footnotesize{1}}}
\multiframe(0,6.5)(10.5,0){1}(50,7){}
\multiframe(0,-1)(10.5,0){1}(40,7){}\put(52,-2){k}
\end{picture}. The set of variables $\xi$, $d\xi$, $d^2\xi$, ...,
$d^{s-1}\xi$ is collectively denoted by $\langle \xi\rangle$. The
traceless parts of the set of constrained variables
$\widehat{\xi}$, $d\widehat{\xi}$, $d^2\widehat{\xi}$, ...,
$d^{s-1}\widehat{\xi}$ is collectively denoted by
\textlquill\,$\widehat{\xi}\,$\textrquill\, and the corresponding
Grassmann algebra by $\widehat{\Xi}$.

\begin{theorem}\label{Hgamma}

For the \underline{unconstrained }g\underline{au}g\underline{e
theor}y, the cohomology of the longitudinal exterior differential
$\gamma$ is the superalgebra freely generated by
\begin{description}
  \item[-] the space-time coordinates $x^\m$,
  \item[-] the curvature tensor $\cal R$ with its partial derivatives,
  \item[-] the ghost $\xi$ and its $s-1$ exterior derivatives,
  \item[-] the antifields $\Phi^*$ with their partial derivatives,
\end{description}
that is,
$$H\Big(\,\gamma\,,\Upsilon_0(\Phi,\Phi^*)\Big)\cong
\Upsilon_0({\cal R},\Phi^*)\otimes \Xi\,.$$ More explicitly,
$$\gamma F\Big(x,[\Phi],[\Phi^*]\Big)=0\quad\Longleftrightarrow \quad F=G\Big(x,[{\cal R}],\langle \xi\rangle,[\Phi^*]\Big)\,+\,\gamma (\ldots)\,,$$
$$G\Big(x,[{\cal R}],\langle \xi\rangle,[\Phi^*]\Big)=\gamma (\ldots) \quad\Longleftrightarrow \quad G=0\,.$$

For the \underline{constrained }g\underline{au}g\underline{e
theor}y, the cohomology of the longitudinal exterior differential
$\gamma$ is the superalgebra freely generated by
\begin{description}
  \item[-] the space-time coordinates $x^\m$,
  \item[-] the traceless components $\widehat{\,[{\cal R}]\,}$ of (the partial
  derivatives of) the curvature tensor,
  \item[-] the Fronsdal tensor $\cal F$ and its partial derivatives,
  \item[-] the traceless components of the ghost $\widehat{\xi}$ and its $s-1$ exterior derivatives,
  \item[-] the antifields $\overline{\Phi}^*$ with their partial derivatives,
\end{description}
that is\footnote{In the notation $\Upsilon_0({\cal R},{\cal
F},\overline{\Phi}^*)$, there is some redundacy in the sense that
the variables $[{\cal R}]$ and $[{\cal F}]$ are not entirely
independent.}
$$H\Big(\,\gamma\,,\Upsilon_0(\overline{\Phi},\overline{\Phi}^*)\Big)\cong
\Upsilon_0({\cal R},{\cal F},\overline{\Phi}^*)\otimes \widehat{
\Xi}\,.$$ More explicitly,
$$\gamma F\Big(x,[\overline{\Phi}],[\overline{\Phi}^*]\Big)=0\quad\Longleftrightarrow \quad
F=G\Big(x,\widehat{\,[{\cal R}]\,},[{\cal F
}],{\mbox{\textlquill}}\,\widehat{\xi}\,{\mbox{\textrquill}},[\overline{\Phi}^*]\Big)\,+\,\gamma
(\ldots)\,,$$
$$G\Big(x,\widehat{\,[{\cal R}]\,},[{\cal F
}],{\mbox{\textlquill}}\,\widehat{\xi}\,{\mbox{\textrquill}},[\overline{\Phi}^*]\Big)=\gamma
(\ldots) \quad\Longleftrightarrow \quad G=0\,.$$
\end{theorem}

Setting the ghosts and antifields to zero, one can make contact
with the standard physical field content. Moreover, by restricting
the gauge-invariant functions to the stationary surfaces defined
by the relations (\ref{Fronsdalequ}) or (\ref{Weylequ}), one may
even put them on-shell in the corresponding approaches. The
results are summarized in the following corollary of Theorem
\ref{Hgamma}:
\begin{corollary}\label{corol}

$\bullet$ A local function of the unconstrained gauge field
$\varphi$ is gauge invariant under unconstrained gauge
transformations if and only if it depends on $\varphi$ via the de
Wit -- Freedman curvature tensor field $\cal R$ and its partial
derivatives. Therefore, on-shell it is a function of the traceless
component of the partial derivatives of the Weyl-like tensor $\cal
W$ only.
$$\delta_\varepsilon f\Big([\varphi]\Big)=0\,,\quad\forall \varepsilon\quad\Longleftrightarrow \quad f=f\Big([{\cal R}]\Big)
\quad\Longleftrightarrow \quad f\approx f(\widehat{\,[{\cal
W}]\,})\,.$$

$\bullet$ A local function of the unconstrained gauge field
$\varphi$ that is invariant under constrained gauge
transformations must depend on the Fronsdal tensor $\cal F$ with all its partial
derivatives, the traceless component of the partial derivatives
of $\cal R$ and on the double trace $\varphi^{\prime\prime}$ of the
gauge field.
$$\delta_{\widehat{\,\varepsilon}} f\Big([\varphi]\Big)=0\,,\quad\forall
\widehat{\varepsilon}\quad\Longleftrightarrow \quad
f=f\Big(\widehat{\,[{\cal R}]\,},[{\cal
F}],[\varphi^{\prime\prime}]\Big)\,.$$

$\bullet$ For constrained gauge fields and parameters, the most
general local gauge-invariant function must depend on $\widetilde{\varphi}$
only through the Fronsdal tensor $\cal F$ with all its partial
derivatives and the traceless component of the partial derivatives
of $\cal R$. Therefore, on-shell it is a function of the traceless
component of the partial derivatives of the Weyl-like tensor $\cal
W$ solely.
$$\delta_{\widehat{\,\varepsilon}} \,g\Big([\widetilde{\varphi}]\Big)=0\,,\quad\forall \widehat{\varepsilon}
\quad\Longleftrightarrow \quad g=g\Big(\widehat{\,[{\cal
R}]\,},[{\cal F}]\Big) \quad\Longleftrightarrow \quad g\approx
f(\widehat{\,[{\cal W}]\,})\,,$$where $f(\widehat{\,[{\cal
R}]\,}):=g(\widehat{\,[{\cal R}]\,},[{\cal F}]=0)$.
\end{corollary}

Some remarks are in order:
\begin{description}
  \item[-] As one can see, the space of on-shell gauge invariant local functions are identical in both
approaches. This confirms the fact that both approaches are
dynamically equivalent, as written in (\ref{dynequiv}). Moreover,
this is in complete agreement with Vasiliev's unfolded formulation
where the variables $\widehat{\,[{\cal W}]\,}$ span the on-shell
twisted adjoint module for the free theory of a rank-$s$ symmetric
tensor gauge field in flat space-time (see Sections 8 and 12 of
\cite{BCIV}).
  \item[-] Corollary \ref{corol} may also be used when writing
down local free ({\it i.e.} linear) field equations that are
gauge-invariant and translation-invariant. Indeed, Corollary
\ref{corol} implies that the most general possibility is an
equation that fixes to zero a local function linear in $[{\cal
R}]$ in the unconstrained case and linear in $[{\cal R}]$ and
$[{\cal F}]$ in the constrained approach. These possibilities
reproduce the ``Weyl cohomology" and ``Einstein cohomology"
obtained through the very different method of \cite{Shaynk} (see
also Section 10 of \cite{BCIV}).
  \item[-] Notice that these algebraic results also hold for local
spacetime functions obtained by the evaluation of the local
function $f$ at an arbitrary field history. Such a spacetime
version of Corollary \ref{corol} is somehow the converse of
deriving the gauge transformations (\ref{gtransfo}) as the most
general transformations leaving invariant the curvature tensor
${\cal R}(x)$ and, in the constrained approach, the Fronsdal
operator ${\cal F}(x)$. (This may be done as a straigthforward
application of the results contained in \cite{DuboisV}.)
\end{description}

\subsection{Local Koszul-Tate cohomology}\label{KTate}

The space $\cal K$ of unconstrained (off-shell) Killing tensor
fields on flat space-time is the vector space of all completely
symmetric tensor fields $\varepsilon_{\m_1\ldots\m_r}(x)$
satisfying (\ref{Killingequ}). The space ${\cal
K}=\bigoplus\limits_{r=0}^\infty{\cal K}_r$ is $\mathbb N$-graded
by the rank $r$ and can be endowed with a structure of $\mathbb
N$-graded commutative algebra via the natural symmetric product
\begin{equation}(\varepsilon_1 \vee\varepsilon_2)_{\m_1\ldots\,\m_{r_1+r_2}}:=
(\varepsilon_1)_{(\,\m_1\ldots\,\m_{r_1}}(\varepsilon_2)_{\m_{r_1+1}\ldots\,\m_{r_1+r_2})}\,.\label{symprod}\end{equation}
One can explicitly check that the rank-$(r_1+r_2)$ symmetric
tensor field $\varepsilon_1 \vee\varepsilon_2$ obeys the off-shell
Killing-like equation. This ensures that the product $\vee$ is
internal in the associative algebra $\cal K$.

\begin{lemma}\label{lem}
For an unconstrained free spin-$s$ ($s\geq 1$) gauge field theory,
the most general off-shell Killing tensor field
$\varepsilon_{\m_1\ldots\m_{s-1}}(x)$ that is a formal power
series in $x$ is a polynomial of degree $s-1$ where the
coefficient of the term of homogeneity degree $t$ is an
irreducible tensor under $\mathfrak{gl}(n)$ characterized by the
Young diagram
\begin{picture}(85,15)(0,2)
\multiframe(10,7.5)(13.5,0){1}(45,7){}\put(60,7.5){$s-1$}
\multiframe(10,0)(13.5,0){1}(25,7){}\put(41,-1){$t$}
\end{picture}. More explicitly,
\bqn
&\partial_{(\m_1}\varepsilon_{\m_2\ldots\m_{s})}(x)=0&\nn\\
&\stackrel{}{\Updownarrow} &\nn\\
&\varepsilon_{\m_1\ldots\,\m_{s-1}}(x)=\sum\limits_{t=0}^{s-1}
\l_{\m_1\ldots\,\m_{s-1}\,,\,\,\n_1\ldots\,\n_t}\,x^{\n_1}\ldots
\,x^{\n_t}\,,\quad\l_{(\m_1\ldots\,\m_{s-1}\,,\,\,\n_1)\n_2\ldots\,\n_t}=0\,.&\label{Killingsol}
\eqn \end{lemma} \noindent This lemma has been derived previously
by several authors in different versions
\cite{Killingtensorsolution,BGST}. In Appendix \ref{shproof}, we
provide a short and new proof of this useful property.

Our main results about Killing tensors (either off-shell or
on-shell) are contained in the following theorem:
\begin{theorem}\label{Killing}
\begin{description}
  \item[(i)]
Let $\{{P}^\m,{M}^{\n\r}\}$ be a basis of the Lie
algebra $\mathfrak{iso}(n-1,1):={\mathbb R}^n\niplus
\mathfrak{so}(n-1,1)$ of the isometry group of the flat space-time
${\mathbb R}^{n-1,1}$, where $\{{P}^\m\}$ is a basis of the
translation algebra ${\mathbb R}^n$ and $\{{M}^{\m\n}\}$ is
a basis of the Lorentz algebra $\mathfrak{so}(n-1,1)$.

The commutative algebra $\cal K$ of unconstrained off-shell
Killing tensor fields is isomorphic to the quotient of the
symmetric algebra $\bigvee(\mathfrak{iso}(n-1,1))$ of the vector
space $\mathfrak{iso}(n-1,1)$ by the relations\begin{equation}
R\,\equiv\, \{\,{P}_{[\m}{M}_{\n\r]}\,,\,
{M}_{[\m\n}{M}_{\r]\s}\,\}\,.\label{rel}\end{equation}
More precisely, any equivalence class of the algebra
$${\cal K}\,\cong\,\bigvee\big(\mathfrak{iso}(n-1,1)\big)\,/\,R$$ can be
represented by a Weyl-ordered polynomial \begin{equation}
S({M},{P})=\sum\limits_{s\geq
1}\sum\limits_{t=0}^{s-1}
\z_{\m_1\ldots\,\m_{s-1}\,,\,\,\n_1\ldots\,\n_t}\,(\,{M}^{\n_1\m_1}\cdots{M}^{\n_t\m_t}
{P}^{\m_{t+1}}\cdots{P}^{\m_{s-1}}+\mbox{perms}\,\,)\,,\label{representatives}\end{equation}
where the coefficients $\z$ are tensors having the same symmetry
properties as the tensors $\l$ in Lemma \ref{lem} and ``perms"
stands for the sum of terms obtained from
${M}^{\n_1\m_1}\cdots{M}^{\n_t\m_t}$
${P}^{\m_{t+1}}\cdots{P}^{\m_{s-1}}$ by performing
all nontrivial permutations of the elements ${P}^\m$ and
${M}^{\m\n}$.
\item[(ii)] All non-trivial on-shell Killing tensor fields
$\widehat{\,\varepsilon\,}_{\m_1\ldots\m_{s-1}}(x,[\varphi])$ (that
are formal power series in $x$) of the constrained theory of
spin-$s$ free gauge field can be represented by the traceless
component of the off-shell Killing tensor fields given in Lemma
\ref{lem}.
\end{description}
\end{theorem}

By looking at the details of Theorems 6.5 and 6.7 of
\cite{Barnich:2000zw} (and their proofs), it is straightforward to
see that the following corollary is equivalent to the point (ii)
of Theorem \ref{Killing}:
\begin{corollary}
\label{Hdeltamod}In the constrained approach, the top forms
$\widehat{\varepsilon}_{\m_1\ldots\m_{s-1}}(x)\,\widehat{\xi}^{*\,\m_1\ldots\m_{s-1}}d^nx$,
where $\widehat{\varepsilon}_{\m_1\ldots\m_{s-1}}(x)$ runs over
all non-trivial on-shell Killing tensor fields, span the local
Koszul-Tate cohomology
$H^n_2\Big(\d\,|\,d\,,\,\Upsilon_0(\overline{\Phi},\overline{\Phi}^*)\Big)$
in top form degree $n$ and in antifield number $2$.
\end{corollary}

Let ${\mathbb R}[\textsc{X},\textsc{P}]$ be the real polynomial
algebra in the variables $\textsc{X}^{\m}$ and $\textsc{P}^{\n}$.
Let us introduce the antisymetric bilinears
$\textsc{M}^{\m\n}:=\textsc{X}^{\m}\textsc{P}^{\n}-\textsc{X}^{\n}\textsc{P}^{\m}$
that provide a realization of the relations (\ref{rel}). The
representative (\ref{representatives}) in the corresponding
realization of $\bigvee(\mathfrak{iso}(n-1,1))/R$ in ${\mathbb
R}[\textsc{X},\textsc{P}]$ is equal to \begin{equation}
\Lambda(\textsc{X},\textsc{P})= \sum\limits_{s\geq
1}\sum\limits_{t=0}^{s-1}
\l_{\m_1\ldots\,\m_{s-1}\,,\,\,\n_1\ldots\,\n_t}\textsc{X}^{\n_1}\ldots
\textsc{X}^{\n_t} \textsc{P}^{\m_1}\ldots\textsc{P}^{\m_{s-1}}\,,
\label{representativesll} \end{equation} where the tensor
$\l_{\m_1\ldots\,\m_{s-1}\,,\,\,\n_1\ldots\,\n_t}$ is a linear
function of the tensor
$\zeta_{\m_1\ldots\,\m_{s-1}\,,\,\,\n_1\ldots\,\n_t}$. They
satisfy the same symmetry properties as the tensors $\l$ in Lemma
\ref{lem}. The isomorphism between the algebra $\cal K$ of
unconstrained off-shell Killing tensor fields (\ref{Killingsol})
and the quotient $\bigvee(\mathfrak{iso}(n-1,1))/R$ is obvious
when the representatives of the quotient are realized as in
(\ref{representativesll}). In that case, the symmetrized tensor
product (\ref{symprod}) in $\cal K$ is mapped to the pointwise
product of polynomials in ${\mathbb R}[\textsc{X},\textsc{P}]$
induced by the following embeddings \begin{equation}{\cal
K}_{s-1}\,\hookrightarrow\, {\mathbb R}[\textsc{X},\textsc{P}]
\,:\, \varepsilon_{\m_1\ldots\,\m_{s-1}}(x) \mapsto
\varepsilon_{\m_1\ldots\,\m_{s-1}}(\textsc{X})\,\textsc{P}^{\m_{1}}
\ldots\textsc{P}^{\m_{s-1}}\,.\label{isomorfism}\end{equation}

The Poincar\'e-Birkhoff-Witt theorem shows that there exists a
canonical isomorphism \textit{of vector spaces} between the
universal enveloping algebra ${\cal U}(\mathfrak{g})$ and the
symmetric algebra $\bigvee(\mathfrak{g})$. Given a unitary
representation of the Poincar\'e group $ISO(n-1,1)$ such that the
generators $\textsc{P}^\m$ and $\textsc{M}^{\n\r}$ are Hermitian
operators acting on some Hilbert space $\{|\phi\rangle\}$, the
corresponding Hermitian operators of the form
(\ref{representatives}) define a unitary representation of the
universal enveloping algebra ${\cal U}(\mathfrak{iso}(n-1,1))$. By
definition, any Lorentz scalar $\textsc{K}(\textsc{P}^2)$ built out of
the quadratic Casimir operator $\textsc{P}^2$ commute with the generators
$\textsc{P}^\m$ and $\textsc{M}^{\n\r}$ of
$\mathfrak{iso}(n-1,1)$. Therefore, the Hermitian operators of the
form (\ref{representatives}) generate global symmetries of the
quadratic action $\langle\phi|\,\textsc{K}|\phi\rangle$ with
kinetic operator $\textsc{K}:=\textsc{K}(\textsc{P}^2)$.

The Weyl algebra $A_{2n}$ is the quotient of the complex
polynomial algebra ${\mathbb C}[\textsc{X},\textsc{P}]$ by the
commutation relations
\begin{equation}\textsc{P}^{\m}\textsc{X}^{\n}-\textsc{X}^{\n}\textsc{P}^{\m}=i\,\eta^{\m\n}\,.
\label{Heisenberg}\end{equation}
A basis of the Lie algebra $\mathfrak{iso}(n-1,1)$ can be realized
by the set $\{\textsc{P}^\m,\textsc{M}^{\n\r}\}\subset A_{2n}$.
The ``off-shell" Minkowski higher-spin Lie algebra ${\tt
hs}_\infty(\mathfrak{iso}(n-1,1))$ is defined by endowing the
realization of ${\cal U}(\mathfrak{iso}(n-1,1))$ in $A_{2n}$ with a Lie algebra
structure by means of the commutator.

If one considers only traceless tensors
$\widehat{\z}$ in the expression (\ref{representatives}) for the
representatives of $\bigvee(\mathfrak{iso}(n-1,1))/R$, then, by
making use of the commutation relation (\ref{Heisenberg}), the
corresponding representatives $S(\textsc{M},\textsc{P})$ can be
rewritten as the normal-ordered polynomial \begin{equation}
H(\textsc{X},\textsc{P}):=\sum\limits_{s\geq
1}\sum\limits_{t=0}^{s-1}
\widehat{\l}_{\m_1\ldots\,\m_{s-1}\,,\,\,\n_1\ldots\,\n_t}\textsc{X}^{\n_1}\ldots\textsc{X}^{\n_t}
\textsc{P}^{\m_1}\ldots\textsc{P}^{\m_{s-1}}\,,\label{representativesl}\end{equation}
where the constant tensors
$\widehat{\l}_{\m_1\ldots\,\m_{s-1}\,,\,\,\n_1\ldots\,\n_t}$
belong to the irreducible $\mathfrak{o}(n-1,1)$-module labeled by
the Young diagram
\begin{picture}(85,15)(0,2)
\multiframe(10,7.5)(13.5,0){1}(45,7){}\put(60,7.5){$s-1$}
\multiframe(10,0)(13.5,0){1}(25,7){}\put(41,-1){$t$}
\end{picture}. The elements (\ref{representativesl}) of the Weyl algebra
are of physical interest because
they commute with any Lorentz scalar $\textsc{K}(\textsc{P}^2)$
built out of $\textsc{P}^2$.

Given a unitary representation of the Weyl algebra $A_{2n}$
defined by (\ref{Heisenberg}) such that $\textsc{X}^\m$ and
$\textsc{P}^\n$ are Hermitian, the Hermitian representatives
(\ref{representativesl}) generate global symmetries of the
quadratic action $\langle\phi|\,\textsc{K}|\phi\rangle$ with
kinetic operator $\textsc{K}=\textsc{K}(\textsc{P}^2)$. For
instance, the Hermitian operators $\textsc{X}^\m:=x^\m$ and
$\textsc{P}^\m:=i\partial^\m$ define a unitary representation of
the Weyl algebra $A_{2n}$ on the functional space of
square-integrable complex bosonic scalar fields $\phi$.
The realization of the enveloping algebra ${\cal U}(\mathfrak{iso}(n-1,1))$ 
in $A_{2n}$ quotiented by the traces admits a faithful unitary
representation acting as the global infinitesimal transformations
\begin{equation}\d_{\widehat{\varepsilon}}\phi\,=\,
i^{s-2}\,\widehat{\varepsilon}_{\n_1\ldots\n_{s-1}}(x)\,\partial^{\n_1}\ldots\partial^{\n_{s-1}}
\phi\,,\label{glinftr}\end{equation} where
$\widehat{\varepsilon}(x)$ is an arbitrary traceless off-shell
Killing tensor field. The higher-order global symmetries
(\ref{glinftr}) should naturally give rise via the Noether theorem
to the list of higher-spin conserved currents explicitly
constructed in \cite{Vasiliev:1999}. It is clear that the
infinitesimal transformations (\ref{glinftr}) are related to the
mappings (\ref{isomorfism}) between Killing tensor fields and
polynomials in $\textsc{X}$ and $\textsc{P}$ in the constrained
approach. As explained in Subsection \ref{Killingsect}, for a
scalar field, the trace in the parameter correspond to global
transformations leaving the scalar field invariant on-shell.
Factoring out the traces of ${\tt
hs}_\infty(\mathfrak{iso}(n-1,1))$ leads to the ``on-shell"
Minkowski higher-spin Lie algebra denoted by ${\tt
hs}(\mathfrak{iso}(n-1,1))$.

Theorem \ref{Killing} implies that, as suggested in the
introduction, the infinite tower of Killing tensors in flat
space-time is related to some Minkowski higher-spin algebra:
\begin{corollary}\label{coro3}
The elements of the off-shell higher-spin algebra ${\tt
hs}_\infty(\mathfrak{iso}(n-1,1))$ are in one-to-one
correspondence with the off-shell Killing tensor fields
$\varepsilon_{\n_1\ldots\n_{s-1}}$ of the unconstrained approach,
while the elements of the on-shell higher-spin algebra ${\tt
hs}(\mathfrak{iso}(n-1,1))$ are in one-to-one correspondence with
the non-trivial on-shell Killing tensor fields
$\widehat{\varepsilon}_{\n_1\ldots\n_{s-1}}$ of the constrained
approach.
\end{corollary}

The bijection is more manifest in the frame-like formulation. The
constructions of the previous higher-spin algebras (together with their relationship
with Killing tensors) are the
analogue of the case of bosonic anti de Sitter / conformal
higher-spin algebras \cite{Mik,Eastwood,Valgebra,Sagnotti:2005} (see
also Section 5 of \cite{BCIV}) except that the Lie algebra
$\mathfrak{iso}(n-1,1)$ must be replaced by
$\mathfrak{so}(n-1,2)$. The former Lie algebra can be obtained
from the latter as an In\"{o}n\"{u}-Wigner contraction in the flat
limit $\Lambda\rightarrow 0$, therefore this also holds for the
corresponding higher-spin algebras. During the redaction of the
present paper, it has been shown by M.A. Vasiliev \cite{Vunf} that
the algebra ${\tt hs}_\infty(\mathfrak{iso}(n-1,1))$ corresponds
to the global symmetries of Minkowski vacuum solution of the
off-shell unfolded equations for the bosonic higher-spin gauge
field theory \cite{Valgebra}. This result confirms the relevance
of our analysis.\\

{\bf Note added:} After having submitted the paper to {\tt arXiv},
we received the work \cite{Nazim} where the on-shell Killing tensors and
the characteristic cohomology groups in form degrees $\leq n-2$
are computed for the constrained spin-$3$ theory; and we have been informed that
slightly more general versions of the point (ii) of Theorem
\ref{Killing} and of Corollary \ref{Hdeltamod} have been obtained
independently by G. Barnich and N. Bouatta \cite{BaBo}.

\section*{Acknowledgements}

We thank L. Gualtieri for early discussions on the spin-$3$ case.


\appendix

\section{Irreducible representations}\label{irreps}

Most of the textbook material reviewed in this section can be
extracted from \cite{Littlewood}. In conformity with the
mathematical literature, we adopt the Euclidean signature
convention.\vspace{1mm}

\subsection{Young diagrams}\label{Young}

Partition of integers play a key role in labeling the irreducible
representations (irreps) of the general linear and orthogonal
groups. The partition of the positive integer $|\l|$ into $r$
integer parts $\l_1$, $\l_2$, ..., $\l_r$ with
$\l_1+\l_2+\ldots+\l_r=|\l|$ and $\l_1\geq \l_2\geq \ldots\geq
\l_r>0$ is denoted by $\l=(\l_1,\l_2, \ldots, \l_r)$. Each such
partition $\l$ specifies a \emph{Young diagram} consisting of
$|\l|$ boxes arranged in $r$ left-adjusted rows, where the length
of the $i$th row is $\l_i$ ($i=1,2,\ldots,r$).

Let $Y$ be the Abelian group made of all formal finite sums of
Young diagrams with integer coefficients. This group is $\mathbb
N$-graded by the number $|\l|$ of boxes: $Y=\sum_{n\in\mathbb N}
Y_n $. The famous ``Littlewood-Richardson rule" defines a
multiplication law which endows $Y$ with a structure of
graded commutative ring.
The product of two Young diagrams $\l$ and $\m$ is
defined as the bilinear mapping to
$$\l\cdot\m=\sum_\nu m_{\l\m\,|\,\n}\,\n\,,$$where the
coefficients $m_{\l\m\,|\,\n}=m_{\m\l\,|\,\n}$ are the number of
distinct labeling of the Young diagram $\n$ obtained from the
following procedure:
\begin{description}
  \item[1.] Label the Young diagram $\m$ by writing the letter
  ``a" in all boxes of its first row, the letter
  ``b" in all boxes of its second row, the letter
  ``c" in all boxes of its third row, {\it etc}.
  \item[2.] Add the labeled boxes of the Young tableau $\m$ to $\l$ in Latin alphabetic order, one letter at a
  time and in such a way that at every stage:
    \begin{description}
    \item[(i)] The resulting diagram is a Young diagram,
    \item[(ii)] No two identical letters appear in the same
    column,
    \item[(iii)] Reading from right to
    left across each row in turn from top to bottom (like in Arabic), the number of
    a's read should always be $\geq$ the number of b's read $\geq$ the number of c's
    read, {\it etc}.
    \end{description}
\end{description}
As one can see, $|\l\cdot\m|=|\l|+|\m|$. A related operation in
$Y$ is the ``division" of $\n$ by $\m$ defined as the bilinear
mapping to
$$\n/\m=\sum_\l m_{\l\,\m\,|\,\n}\,\l\,,$$where the sum is over Young diagrams $\l$
such that the product $\l\cdot\m$ contains the term $\n$ (with
coefficient $m_{\l\m\,|\,\n}$).

The following obvious lemma will be used many times to simplify
the computation of the cohomology of the differential $\g$ along
the gauge orbits:
\begin{lemma}\label{twocolYd}
Let $m$ and $n$ be two strictly positive integers such that $m\geq
n$.

The product of two rows of respective lengths $m$ and $n$ is the
sum of the two-row Young diagram obtained by putting the shortest
row on the bottom of the longest and the product of two rows of
respective lengths $m+1$ and $n-1$.

$$
\begin{picture}(130,15)(-5,2)
\put(25,12){$m$}\multiframe(0,0)(10.5,0){1}(60,10){}
\put(70,1){$\cdot$}
\put(95,12){$n$}\multiframe(80,0)(10.5,0){1}(40,10){}
\end{picture}
=
\begin{picture}(75,15)(-5,2)
\put(25,17){$m$} \multiframe(0,5)(10.5,0){1}(60,10){}
\multiframe(0,-5)(10.5,0){1}(40,10){} \put(16,-13){$n$}
\end{picture}
+
\begin{picture}(120,15)(-5,2)
\put(20,12){$m+1$}\multiframe(0,0)(10.5,0){1}(65,10){}
\put(75,1){$\cdot$}
\put(90,12){$n-1$}\multiframe(85,0)(10.5,0){1}(35,10){}
\end{picture}
$$
\end{lemma}

\subsection{Kronecker products and branching rules}\label{Kronecker}

Irreps of $\mathfrak{gl}(n)$ may be labeled by $\{\l\}$ where the
partition $\l$ serves to specify the symmetry properties of the
corresponding rank-$|\,\l|$ covariant (or contravariant) tensors
forming the basis of this irreducible Lie algebra module. Let
$Y_+$ be the Abelian monoid made of all formal finite sums of
Young diagrams with non-negative integer coefficients. Finite
direct sums of irreps of $\mathfrak{gl}(n)$ may therefore be
labeled by $Y_+$ via the
rule$$\{\,m\,\m\,+\,n\,\n\,\}\,=\,m\,\{\m\}\,\oplus\,
n\,\{\n\}\,,$$ where the positive integer coefficients
$m,n\in\mathbb N$ must be interpreted as the multiplicity of the
corresponding representation.

The evaluation of the Kronecker product of two
$\mathfrak{gl}(n)$-irreps $\{\l\}$ and $\{\m\}$ can be done by
means of the Littlewood-Richardson rule which gives
\begin{equation}\{\l\}\otimes\{\m\}=\{\l\cdot\m\}=\bigoplus_\nu\,
m_{\l\m\,|\,\n}\{\n\}\,.\label{LRgl}\end{equation} A related
operation is that of contraction of one set of contravariant
indices of symmetry $\m$ with a subset of a set of covariant
tensor indices of symmetry $\n$ to yield a sum of covariant
tensors with indices of symmetry $\l$ given by the division
rule$$\{\n\}/\{\m\}=\{\n/\m\}=\bigoplus_\l
\,m_{\l\m\,|\,\n}\,\{\l\}\,.$$

The irreps of $\mathfrak{gl}(n)$ may be reduced to irreps of
$\mathfrak{o}(n)$ by extracting all possible trace terms formed by
contraction with the metric tensor $\eta$ and its inverse. The
corresponding irreps are labeled by $[\s]$. The reduction is given
by the branching rule\begin{equation}\mathfrak{gl}(n)\downarrow
\mathfrak{o}(n)\quad:\quad \{\l\}\downarrow
[\l/\Delta]\,,\label{reduction}\end{equation} where $\Delta$ is
the formal infinite sum$$\Delta=\,1\,+\begin{picture}(30,15)(-5,2)
\multiframe(0,0)(10.5,0){2}(10,10){}{}
\end{picture}
+
\begin{picture}(55,15)(-5,2)
\multiframe(0,0)(10.5,0){4}(10,10){}{}{}{}
\end{picture}
+
\begin{picture}(30,15)(-5,2)
\multiframe(0,5)(10.5,0){2}(10,10){}{}
\multiframe(0,-5)(10.5,0){2}(10,10){}{}
\end{picture}
+
\begin{picture}(80,15)(-5,2)
\multiframe(0,0)(10.5,0){6}(10,10){}{}{}{}{}{}
\end{picture}
+
\begin{picture}(55,15)(-5,2)
\multiframe(0,5)(10.5,0){4}(10,10){}{}{}{}
\multiframe(0,-5)(10.5,0){2}(10,10){}{}
\end{picture}
+
\begin{picture}(30,15)(-5,2)
\multiframe(0,10)(10.5,0){2}(10,10){}{}
\multiframe(0,0)(10.5,0){2}(10,10){}{}
\multiframe(0,-10)(10.5,0){2}(10,10){}{}
\end{picture}
+\ldots
$$ corresponding to the sum of all possible plethysms of the metric tensor.
The decomposition (\ref{reduction}) actually has a useful converse
\begin{equation}\mathfrak{o}(n)\uparrow \mathfrak{gl}(n)\quad:\quad
[\l]\uparrow \{\l/\Delta^{-1}\}\,,\label{extension}\end{equation}
because the series $\Delta$ has an inverse
$$\Delta^{-1}=\,1\,-\begin{picture}(30,15)(-5,2)
\multiframe(0,0)(10.5,0){2}(10,10){}{}
\end{picture}
+
\begin{picture}(43,15)(-5,2)
\multiframe(0,5)(10.5,0){3}(10,10){}{}{}
\multiframe(0,-5)(10.5,0){1}(10,10){}
\end{picture}
-
\begin{picture}(55,15)(-5,2)
\multiframe(0,10)(10.5,0){4}(10,10){}{}{}{}
\multiframe(0,0)(10.5,0){1}(10,10){}
\multiframe(0,-10)(10.5,0){1}(10,10){}
\end{picture}
-
\begin{picture}(43,15)(-5,2)
\multiframe(0,5)(10.5,0){3}(10,10){}{}{}
\multiframe(0,-5)(10.5,0){3}(10,10){}{}{}
\end{picture}
+\ldots
$$(The material reviewed in
this paragraph finds its origin in the book \cite{King}.)

The operation (\ref{reduction}) leads to a formal finite sum of
irreps, some of which with strictly negative integer coefficients
that have to be interpreted as constraints on some trace of the
corresponding tensor basis. (Remark: These constraints are not
preserved by the full $\mathfrak{gl}(n)$ algebra.) We introduce
the notation
$$\{\l-\m\}\equiv\{\l\}\ominus\{\m\}\,,$$
for later convenience.

\section{Proof of Theorem \ref{Hgamma}}\label{proof1}

Our proof makes a decisive use of the simple fact that
contractible pairs drop out from the (co)homology. The plan of our
proof is therefore very simple: provide a new set of generators
for which the contractible pairs are manifest. This convenient set
of generators is identified via the decomposition of the jet
bundle in irreducible modules of either $\mathfrak{gl}(n)$ or
$\mathfrak{o}(n)$ algebras. This strategy follows the lines of previous
computations of $H^0(\g)$ for other gauge theories, such as
completely antisymmetric \cite{Henneaux:1998rp}, spin-two
\cite{Boulanger:2000rq} and two-column mixed symmetry \cite{mixed}
gauge fields.\vspace{1mm}

\subsection{Contractible pairs}

Let $\cal A$ be the supercommutative differential algebra over the
field $\mathbb K$ that is (i) freely generated by the variables
$x^i$, $y^a$ and $z^a$ whose respective Grassmann parities are
related by $$\epsilon(z^a)=\epsilon(y^a)+1\,,$$ and (ii) endowed
with the differential $\Delta$ defined via
$$\Delta x^i=0\,,\quad \Delta y^a=z^a\,,\quad\Delta z^a=0\,,$$ and the Leibnitz rule.
The differential superalgebra ${\cal A}$ is graded by the
degree of homogeneity in the variables $y^a$ for which the
differential $\Delta$ is of degree minus one. The pairs
$(y^a,z^a)$ are called {\textit{contractible pairs}}. This
terminology follows from the well-known lemma\footnote{For a
proof, see {\it e.g.} Section 8.3.2 of \cite{Henneauxbook}.}:
\begin{lemma}\label{contractpair}
The differential superalgebra $({\cal A},\Delta)$ provides a
homological resolution of the polynomial algebra ${\mathbb
K}[x^i]$. More precisely, the homology $H(\Delta,{\cal A})$
decomposes according to the degree of homogeneity in the variables
$y^a$ as follows:
$$H_0(\Delta,{\cal A})= {\mathbb K}[x^i]\,,\quad H_k(\Delta,{\cal
A})=0\,,\quad k\neq 0\,.$$
\end{lemma}

\subsection{Computation of $H(\gamma)$}

The lemma \ref{twocolYd} combined with the rule (\ref{LRgl}) leads
to the following rule for the Kronecker product
\begin{equation}\label{Krule}
\{\begin{picture}(60,15)(-5,2) \put(25,11){r}
\multiframe(0,2)(10.5,0){1}(50,7){}
\end{picture}\}\otimes\{\begin{picture}(50,15)(-5,2) \put(20,11){s}
\multiframe(0,2)(10.5,0){1}(40,7){}
\end{picture}\}
= \{\begin{picture}(60,15)(-5,2)\put(9,17){\tiny{max(r,s)}}
\multiframe(0,6.5)(10.5,0){1}(50,7){}
\multiframe(0,-1)(10.5,0){1}(40,7){}\put(5,-9){\tiny{min(r,s)}}
\end{picture}
\}\,\oplus\,\Big(\,\{\begin{picture}(70,15)(-5,2)
\put(10,12){\tiny{max(r,s)+1}} \multiframe(0,2)(10.5,0){1}(60,7){}
\end{picture}\}\otimes\{\begin{picture}(40,15)(-5,2) \put(0,12){\tiny{min(r,s)-1}}
\multiframe(0,2)(10.5,0){1}(30,7){}
\end{picture}\}\,\Big)\end{equation}
that will be very useful in the sequel.

\subsubsection{Unconstrained case}

The set of generators are the variables $[\Phi]$. The variables
$[\varphi]$ and $[\xi]$ must be compared in the
$\mathfrak{gl}(n)$-module of covariant tensors of the same rank.
The covariant tensors $\partial^k\varphi$ belonging to the
Kronecker product $\{\begin{picture}(60,15)(-5,2) \put(25,10){k}
\multiframe(0,2)(10.5,0){1}(50,7){}
\end{picture}\}\otimes\{\begin{picture}(50,15)(-5,2) \put(20,10){s}
\multiframe(0,2)(10.5,0){1}(40,7){}
\end{picture}\}$
must be compared with the covariant tensors $\partial^{k+1}\xi$
belonging to the $\mathfrak{gl}(n)$-module labeled by
$\{\begin{picture}(70,15)(-5,2) \put(24,10){k+1}
\multiframe(0,2)(10.5,0){1}(60,7){}
\end{picture}\}\otimes\{\begin{picture}(40,15)(-5,2) \put(10,10){s-1}
\multiframe(0,2)(10.5,0){1}(30,7){}
\end{picture}\}$. When doing this comparison, we remember that
they are related via the formula (\ref{gammaformula}) and look for
a decomposition of the variables $[\Phi]$ into a set
$\{x^i,y^a,z^a\}$ in analogy with Lemma
\ref{contractpair}.\vspace{1mm}

Two distinct cases have to be addressed:
\begin{description}
  \item[$\diamond$] $\underline{-1\leq k\leq s-2}$: This means that $0\leq k+1\leq
  s-1$, therefore the covariant tensors $\partial^{k+1}\xi$
  decomposes as the direct sum
\begin{equation}\label{k+1.s-1}
\{\begin{picture}(50,15)(-5,2) \put(12,11){k+1}
\multiframe(0,2)(10.5,0){1}(40,7){}
\end{picture}\}\otimes\{\begin{picture}(60,15)(-5,2) \put(18,11){s-1}
\multiframe(0,2)(10.5,0){1}(50,7){}
\end{picture}\}
= \{\begin{picture}(60,15)(-5,2)\put(15,17){s-1}
\multiframe(0,6.5)(10.5,0){1}(50,7){}
\multiframe(0,-1)(10.5,0){1}(40,7){}\put(10,-11){k+1}
\end{picture}
\}\,\oplus\,\Big(\,\{\begin{picture}(40,15)(-5,2) \put(14,12){k}
\multiframe(0,2)(10.5,0){1}(30,7){}
\end{picture}\}\otimes\{\begin{picture}(70,15)(-5,2) \put(28,12){s}
\multiframe(0,2)(10.5,0){1}(60,7){}
\end{picture}\}\,\Big)\,,
\end{equation}
where the second term on the r.h.s. is absent for $k=-1$. The
first term corresponds to irreducible covariant tensors
$d^{k+1}\xi$ ($k+1\leq s-1$) which are $x^i$ variables for the
longitudinal differential $\g$ according to the notation of
Lemma \ref{contractpair}. The second term on the r.h.s. of
(\ref{k+1.s-1}) corresponds to covariant tensors $z^a$ in the
Kronecker product $\{\begin{picture}(40,15)(-5,2) \put(14,12){k}
\multiframe(0,2)(10.5,0){1}(30,7){}
\end{picture}\}\otimes\{\begin{picture}(50,15)(-5,2) \put(18,12){s}
\multiframe(0,2)(10.5,0){1}(40,7){}
\end{picture}\}$ that form contractible pairs with the
variables $y^a\equiv \partial^k\varphi$ for $k\geq 0$.
  \item[$\diamond$] $\underline{k\geq s-1}$: The Kronecker product corresponding to $\partial^k\varphi$
  decomposes as\begin{equation}\label{k.s}
\{\begin{picture}(60,15)(-5,2) \put(25,11){k}
\multiframe(0,2)(10.5,0){1}(50,7){}
\end{picture}\}\otimes\{\begin{picture}(50,15)(-5,2) \put(20,11){s}
\multiframe(0,2)(10.5,0){1}(40,7){}
\end{picture}\}
= \{\begin{picture}(60,15)(-5,2)\put(18,17){k}
\multiframe(0,6.5)(10.5,0){1}(50,7){}
\multiframe(0,-1)(10.5,0){1}(40,7){}\put(18,-10){s}
\end{picture}
\}\,\oplus\,\Big(\,\{\begin{picture}(70,15)(-5,2) \put(20,12){k+1}
\multiframe(0,2)(10.5,0){1}(60,7){}
\end{picture}\}\otimes\{\begin{picture}(40,15)(-5,2) \put(8,12){s-1}
\multiframe(0,2)(10.5,0){1}(30,7){}
\end{picture}\}\,\Big)\end{equation}
The first term on the r.h.s. corresponds to $(k-s)$th partial
derivatives of the de Wit--Freedman tensor $\partial^{k-s}{\cal
R}\equiv x^j$. (The Bianchi identities are responsible for the
fact that the partial derivatives of the linearized curvature
tensor belongs to the irrep. labeled by two-row Young diagrams.)
Eventually, the second term on the r.h.s. of (\ref{k.s})
correspond to variables $y^b$ forming a contractible pair together with
the variables $z^b\equiv
\partial^{k+1}\xi$.
\end{description}

The application of Lemma \ref{contractpair} is straightforward
since the pairs have been explicitly separated. (The antifield
variables $[\Phi^*]$ are inert under the action of the exterior
differential.) This leads to the first part of Theorem
\ref{Hgamma}.

\subsubsection{Constrained case: change of basis}

The basis we start with corresponds to $[\overline{\Phi}]$. The
basis elements $[\widetilde{\varphi}]$ and $[\widehat{\xi}]$
must be compared in $\mathfrak{gl}(n)$-modules of covariant
tensors of the same rank. To do so, we have first to apply the
``converse" branching rule (\ref{extension}) since the fields
$\overline{\Phi}$ satisfy some trace constraints. Firstly,
$\widetilde{\varphi}$ are double-traceless covariant tensors of
rank $s$, thus they span the direct sum
$[\begin{picture}(60,15)(-5,2) \put(25,10){s}
\multiframe(0,2)(10.5,0){1}(50,7){}
\end{picture}]\oplus[\begin{picture}(50,15)(-5,2) \put(15,10){s-2}
\multiframe(0,2)(10.5,0){1}(40,7){}
\end{picture}]$ of irreducible $\mathfrak{o}(n)$-modules, which
may also be labeled by $\{\begin{picture}(60,15)(-5,2)
\put(25,10){s} \multiframe(0,2)(10.5,0){1}(50,7){}
\end{picture}\}\ominus\{\begin{picture}(30,15)(-5,2) \put(4,10){s-4}
\multiframe(0,2)(10.5,0){1}(20,7){}
\end{picture}\}$. Secondly,
$\widehat{\xi}$ are traceless covariant tensors of rank $s-1$,
thus they span the irreducible $\mathfrak{o}(n)$-module labeled by
$[\begin{picture}(50,15)(-5,2) \put(15,10){s-1}
\multiframe(0,2)(10.5,0){1}(40,7){}
\end{picture}]\,\,\uparrow\,\{\begin{picture}(50,15)(-5,2) \put(15,10){s-1}
\multiframe(0,2)(10.5,0){1}(40,7){}
\end{picture}\}\ominus\{\begin{picture}(40,15)(-5,2) \put(10,10){s-3}
\multiframe(0,2)(10.5,0){1}(30,7){}
\end{picture}\}$.\vspace{2mm}

Consequently, the comparison between the tensors
$\partial^k\widetilde{\varphi}$ and $\partial^{k+1}\widehat{\xi}$
is more involved in the constrained approach. Actually, it is
practical to consider the gauge fields as unconstrained and impose
the constraints separately. More specifically, when $k\geq 2$, it
turns out to be convenient (and of physical significance) to split
the basis elements $\partial^k\widetilde{\varphi}$ into the tensors
$\widehat{\partial^k}\varphi$ and $\partial^{k-2}{\cal F}$ subject
to the constraints $\widehat{\partial^k}\varphi^{\prime\prime}=0$.
Let us now explain what is meant by this splitting and how it is
performed.
\begin{description}
  \item[-] To start with, we consider the product of $k$ partial
derivatives $\partial_{\m_1}\ldots\partial_{\m_k}$ as covariant
tensor in the irreducible $\mathfrak{gl}(n)$-module labeled by
$\{\begin{picture}(60,15)(-5,2) \put(25,10){k}
\multiframe(0,2)(10.5,0){1}(50,7){}
\end{picture}\}$ which can be reduced to the sum of its traceless
part $\widehat{\partial^k}$ labeled by
$\{\begin{picture}(60,15)(-5,2) \put(25,10){k}
\multiframe(0,2)(10.5,0){1}(50,7){}
\end{picture}\}\ominus\{\begin{picture}(40,15)(-5,2) \put(8,10){k-2}
\multiframe(0,2)(10.5,0){1}(30,7){}
\end{picture}\}\,\downarrow\,
[\begin{picture}(60,15)(-5,2) \put(25,10){k}
\multiframe(0,2)(10.5,0){1}(50,7){}
\end{picture}]$ and a part
proportional to its trace $\Box\partial^{k-2}$ labeled by
$\{\begin{picture}(40,15)(-5,2) \put(8,10){k-2}
\multiframe(0,2)(10.5,0){1}(30,7){}
\end{picture}\}$.

  \item[-] The space spanned by the tensors
$\partial^k\widetilde{\varphi}$ is by definition exactly the same
as the space spanned by the tensors $\partial^k\varphi$ quotiented
by the space spanned by $\partial^k\varphi^{\prime\prime}$ that is
labeled by the Kronecker product $\{\begin{picture}(60,15)(-5,2)
\put(25,10){k} \multiframe(0,2)(10.5,0){1}(50,7){}
\end{picture}\}\otimes\{\begin{picture}(35,15)(-5,2) \put(7,10){s-4}
\multiframe(0,2)(10.5,0){1}(25,7){}
\end{picture}\}$.

  \item[-] Then we perform the following invertible change of
basis $\{\partial^k\varphi\}\longleftrightarrow
\{\widehat{\partial^k}\varphi\,,\,\partial^{k-2}\Box\varphi\}$.
Moreover, it is possible to perform a triangular invertible linear
change of
variables $\Box\varphi\longleftrightarrow {\cal F}$ relating the
Laplacian of the gauge field to the Fronsdal tensor
(\ref{Fronsdaltensor}).

  \item[-] Putting all these remarks together, we have proved that
the span of $\{\partial^k\widetilde{\varphi}\}$ is the quotient of
the span $\{\widehat{\partial^k}\varphi\,,\,\partial^{k-2}{\cal
F}\}$ divided by the span of
$\{\partial^k\varphi^{\prime\prime}\}$.
\end{description}

To summarize, we must compare the following four Kronecker
products\bqn\widehat{\partial^k}\varphi&:&\quad\Big(\{\begin{picture}(60,15)(-5,2)
\put(25,10){k} \multiframe(0,2)(10.5,0){1}(50,7){}
\end{picture}\}\ominus\{\begin{picture}(40,15)(-5,2) \put(8,10){k-2}
\multiframe(0,2)(10.5,0){1}(30,7){}
\end{picture}\}\Big)\otimes
\{\begin{picture}(60,15)(-5,2) \put(25,10){s}
\multiframe(0,2)(10.5,0){1}(50,7){}
\end{picture}\}\label{k-2hatvarphi}\\
\partial^k\varphi^{\prime\prime}&:&\quad\quad\quad\quad\quad\quad\quad\quad\{\begin{picture}(60,15)(-5,2)
\put(25,10){k} \multiframe(0,2)(10.5,0){1}(50,7){}
\end{picture}\}\otimes\{\begin{picture}(30,15)(-5,2) \put(4,10){s-4}
\multiframe(0,2)(10.5,0){1}(20,7){}
\end{picture}\}
\label{kphidoubletrace}\\
\partial^{k-2}{\cal
F}&:&\quad\quad\quad\quad\,\,\,\,\quad\quad\quad\quad\quad\{\begin{picture}(40,15)(-5,2)
\put(8,10){k-2} \multiframe(0,2)(10.5,0){1}(30,7){}
\end{picture}\}\otimes\{\begin{picture}(60,15)(-5,2)
\put(25,10){s} \multiframe(0,2)(10.5,0){1}(50,7){}
\end{picture}\}\label{k-2F}\\
\partial^{k+1}\widehat{\xi}&:&\,\quad\quad\quad\quad\quad\quad\quad\{\begin{picture}(70,15)(-5,2)
\put(25,10){k+1} \multiframe(0,2)(10.5,0){1}(60,7){}
\end{picture}\}\otimes\Big(\{\begin{picture}(50,15)(-5,2) \put(15,10){s-1}
\multiframe(0,2)(10.5,0){1}(40,7){}
\end{picture}\}\ominus\{\begin{picture}(40,15)(-5,2) \put(10,10){s-3}
\multiframe(0,2)(10.5,0){1}(30,7){}
\put(50,5){.}
\end{picture}\}\Big)\label{k+1C}
\eqn

\subsubsection{Constrained case: decomposition of Kronecker products}

Three cases have to be distinguished:
\begin{description}
  \item[$\ast$] $\underline{-1\leq k\leq s-4}$:
The Kronecker product corresponding to the $(k+1)$th partial
derivative of the trace constraint on the ghost $\xi$ can be
reexpressed as the direct sum \bqn\label{k+1.s-3}
\{\begin{picture}(40,15)(-5,2) \put(6,10){k+1}
\multiframe(0,2)(10.5,0){1}(30,7){}
\end{picture}\}
\otimes
\{\begin{picture}(50,15)(-5,2) \put(16,10){s-3}
\multiframe(0,2)(10.5,0){1}(40,7){}
\end{picture}\}
&=& \{\begin{picture}(50,15)(-5,2)\put(12,16){s-3}
\multiframe(0,6.5)(10.5,0){1}(40,7){}
\multiframe(0,-1)(10.5,0){1}(30,7){}\put(6,-11){k+1}
\end{picture}\}
\oplus \{\begin{picture}(55,15)(-5,2)\put(12,16){s-2}
\multiframe(0,6.5)(10.5,0){1}(45,7){}
\multiframe(0,-1)(10.5,0){1}(25,7){}\put(12,-11){k}
\end{picture}\}
\oplus \{\begin{picture}(60,15)(-5,2)\put(15,16){s-1}
\multiframe(0,6.5)(10.5,0){1}(50,7){}
\multiframe(0,-1)(10.5,0){1}(20,7){}\put(3,-11){k-1}
\end{picture}\}
\,\nn\\\nn\\ &&\oplus
\Big(\,\{\begin{picture}(30,15)(-5,2)\put(3,11){k-2}
\multiframe(0,2)(10.5,0){1}(20,7){}
\end{picture}\}
\otimes \{\begin{picture}(70,15)(-5,2) \put(30,11){s}
\multiframe(0,2)(10.5,0){1}(60,7){}
\end{picture}\} \,\Big)\,, \eqn
where this equation is obtained by three successive application of
the rule (\ref{Krule}) when $k\geq 2$. (If $k<2$ then the formula
(\ref{k+1.s-3}) remains valid but the terms that would not be
well-defined are absent.)

Combining (\ref{k+1.s-1}) with (\ref{k+1.s-3}) leads to the
following decomposition of the Kronecker product (\ref{k+1C}):
\bqn &&
 \{\begin{picture}(70,15)(-5,2)\put(23,16){s-1}
\multiframe(0,6.5)(10.5,0){1}(60,7){}
\multiframe(0,-1)(10.5,0){1}(30,7){}\put(6,-11){k+1}
\end{picture}\}
\ominus \{\begin{picture}(50,15)(-5,2)\put(12,16){s-3}
\multiframe(0,6.5)(10.5,0){1}(40,7){}
\multiframe(0,-1)(10.5,0){1}(30,7){}\put(6,-11){k+1}
\end{picture}\}
\ominus \{\begin{picture}(55,15)(-5,2)\put(12,16){s-2}
\multiframe(0,6.5)(10.5,0){1}(45,7){}
\multiframe(0,-1)(10.5,0){1}(25,7){}\put(12,-11){k}
\end{picture}\}
\ominus \{\begin{picture}(60,15)(-5,2)\put(15,16){s-1}
\multiframe(0,6.5)(10.5,0){1}(50,7){}
\multiframe(0,-1)(10.5,0){1}(20,7){}\put(3,-11){k-1}
\end{picture}\}
\nn\\\nn\\ &&\quad\oplus\,\,
\Big(\,(\,\,\{\begin{picture}(35,15)(-5,2)\put(12,11){k}
\multiframe(0,2)(10.5,0){1}(25,7){}
\end{picture}\}\ominus\{\begin{picture}(30,15)(-5,2)\put(3,11){k-2}
\multiframe(0,2)(10.5,0){1}(20,7){}
\end{picture}\}\,\,)
\,\otimes\, \{\begin{picture}(70,15)(-5,2) \put(30,10){s}
\multiframe(0,2)(10.5,0){1}(60,7){}
\end{picture}\}\,\Big)\,. \label{complement}\eqn
Therefore the span of tensors $\partial^{k+1}\widehat{\xi}$ when
$k+1$ ranges from $0$ till $s-1$ contains the irreducible
$\mathfrak{o}(n)$-modules labeled by
$$
\{\begin{picture}(70,15)(-5,2)\put(23,16){s-1}
\multiframe(0,6.5)(10.5,0){1}(60,7){}
\multiframe(0,-1)(10.5,0){1}(30,7){}\put(6,-11){k+1}
\end{picture}\}
\ominus \{\begin{picture}(50,15)(-5,2)\put(12,16){s-3}
\multiframe(0,6.5)(10.5,0){1}(40,7){}
\multiframe(0,-1)(10.5,0){1}(30,7){}\put(6,-11){k+1}
\end{picture}\}
\ominus \{\begin{picture}(55,15)(-5,2)\put(12,16){s-2}
\multiframe(0,6.5)(10.5,0){1}(45,7){}
\multiframe(0,-1)(10.5,0){1}(25,7){}\put(12,-11){k}
\end{picture}\}
\ominus \{\begin{picture}(60,15)(-5,2)\put(15,16){s-1}
\multiframe(0,6.5)(10.5,0){1}(50,7){}
\multiframe(0,-1)(10.5,0){1}(20,7){}\put(3,-11){k-1}
\end{picture}\}\quad\uparrow\quad[\begin{picture}(70,15)(-5,2)\put(23,16){s-1}
\multiframe(0,6.5)(10.5,0){1}(60,7){}
\multiframe(0,-1)(10.5,0){1}(30,7){}\put(6,-11){k+1}
\end{picture}]$$
corresponding to the set of variables ${\mbox{\textlquill}}\,\widehat{\xi}\,{\mbox{\textrquill}}$
introduced in Subsection \ref{Th1}. The complement of the space spanned by the variables ${\mbox{\textlquill}}\,\widehat{\xi}\,{\mbox{\textrquill}}$
is the module labeled by the diagram in the second line of (\ref{complement}) the
basis elements of which form contractible pairs together with the
tensors $\widehat{\partial^k}\varphi$ described in
(\ref{k-2hatvarphi}).

The basis elements of the jet space that remain are therefore in
the cohomology since they cannot be associated with any derivative
of ghost. They are the derivatives $\partial^{k-2}\cal F$ (\ref{k-2F}) of the
Fronsdal tensor subject to the constraints
$\partial^k\varphi^{\prime\prime}=0$ (\ref{kphidoubletrace}).

  \item[$\ast$] $\underline{s-3\leq k\leq s+1}$:
This case contains five subcases each of which must be treated
separately. The final result reproduces the corresponding general
results of the two other cases, so we leave the explicit check as
an exercise for the reader.

  \item[$\ast$] $\underline{k\geq s+2}$: This is equivalent to $k-2\geq s$ therefore
  we apply the rule
 (\ref{Krule}) three times on the Kronecker product (\ref{k-2F}),
associated with the derivatives $\partial^{k-2}\cal F$ of the
Fronsdal operator, and get as a result \bqn\label{k-2.s}
\{\begin{picture}(50,15)(-5,2) \put(14,10){k-2}
\multiframe(0,2)(10.5,0){1}(40,7){}
\end{picture}\}
\otimes
 \{\begin{picture}(40,15)(-5,2) \put(14,10){s}
\multiframe(0,2)(10.5,0){1}(30,7){}
\end{picture}\}
&=& \{\begin{picture}(50,15)(-5,2)\put(12,16){k-2}
\multiframe(0,6.5)(10.5,0){1}(40,7){}
\multiframe(0,-1)(10.5,0){1}(30,7){}\put(13,-11){s}
\end{picture}\}
\oplus \{\begin{picture}(55,15)(-5,2)\put(13,16){k-1}
\multiframe(0,6.5)(10.5,0){1}(45,7){}
\multiframe(0,-1)(10.5,0){1}(25,7){}\put(6,-11){s-1}
\end{picture}\}
\oplus \{\begin{picture}(60,15)(-5,2)\put(18,16){k}
\multiframe(0,6.5)(10.5,0){1}(50,7){}
\multiframe(0,-1)(10.5,0){1}(20,7){}\put(3,-11){s-2}
\end{picture}\}
\,\nn\\\nn\\ &&\oplus \Big(\, \{\begin{picture}(70,15)(-5,2)
\put(20,11){k+1} \multiframe(0,2)(10.5,0){1}(60,7){}
\end{picture}\}
\otimes\{\begin{picture}(30,15)(-5,2)\put(3,11){s-3}
\multiframe(0,2)(10.5,0){1}(20,7){}
\end{picture}\} \,\Big)\,. \eqn
The Damour-Deser identity relates the trace of the curvature
tensor to $s-2$ curls of the Fronsdal tensor ${\cal
R}^\prime\propto d^{s-2}\cal F$ \cite{DD,BB2}.
Consequently, the right-hand-side of the first line of
(\ref{k-2.s}) can be expressed entirely in terms of derivatives of the trace ${\cal R}^\prime$ of the Riemann tensor. Combining (\ref{k-2.s})
with (\ref{k.s}) we get that the $k$th derivatives of the
double-traceless gauge field of rank $s$ decompose as follows \bqn
&&\{\begin{picture}(60,15)(-5,2) \put(23,10){k}
\multiframe(0,2)(10.5,0){1}(50,7){}
\end{picture}\}
\otimes\Big(
\{\begin{picture}(40,15)(-5,2) \put(14,10){s}
\multiframe(0,2)(10.5,0){1}(30,7){}
\end{picture}\}
\ominus \{\begin{picture}(30,15)(-5,2) \put(4,10){s-4}
\multiframe(0,2)(10.5,0){1}(20,7){}
\end{picture}\}\Big)
\nn\\\nn\\
&&= \{\begin{picture}(70,15)(-5,2)\put(28,16){k}
\multiframe(0,6.5)(10.5,0){1}(60,7){}
\multiframe(0,-1)(10.5,0){1}(30,7){}\put(13,-11){s}
\end{picture}\}
\ominus \{\begin{picture}(50,15)(-5,2)\put(12,16){k-2}
\multiframe(0,6.5)(10.5,0){1}(40,7){}
\multiframe(0,-1)(10.5,0){1}(30,7){}\put(13,-11){s}
\end{picture}\}
\ominus \{\begin{picture}(55,15)(-5,2)\put(13,16){k-1}
\multiframe(0,6.5)(10.5,0){1}(45,7){}
\multiframe(0,-1)(10.5,0){1}(25,7){}\put(6,-11){s-1}
\end{picture}\}
\ominus \{\begin{picture}(60,15)(-5,2)\put(18,16){k}
\multiframe(0,6.5)(10.5,0){1}(50,7){}
\multiframe(0,-1)(10.5,0){1}(20,7){}\put(3,-11){s-2}
\end{picture}\}\nn\\\nn\\
&&\quad\oplus \Big(\{\begin{picture}(50,15)(-5,2) \put(14,10){k-2}
\multiframe(0,2)(10.5,0){1}(40,7){}
\end{picture}\}
\otimes
 \{\begin{picture}(40,15)(-5,2) \put(14,10){s}
\multiframe(0,2)(10.5,0){1}(30,7){}
\end{picture}\}\Big)\,\,\ominus\,\, \Big(\{\begin{picture}(60,15)(-5,2)
\put(25,10){k} \multiframe(0,2)(10.5,0){1}(50,7){}
\end{picture}\}\otimes\{\begin{picture}(30,15)(-5,2) \put(4,10){s-4}
\multiframe(0,2)(10.5,0){1}(20,7){}
\end{picture}\}\Big)\nn\\\nn\\
&&\quad\oplus\Big(\{\begin{picture}(70,15)(-5,2) \put(25,10){k+1}
\multiframe(0,2)(10.5,0){1}(60,7){}
\end{picture}\}\otimes(\,\,\{\begin{picture}(50,15)(-5,2) \put(15,10){s-1}
\multiframe(0,2)(10.5,0){1}(40,7){}
\end{picture}\}\ominus\{\begin{picture}(40,15)(-5,2) \put(10,10){s-3}
\multiframe(0,2)(10.5,0){1}(30,7){}
\end{picture}\}\,\,)\,\Big)
\label{complements}\eqn
\end{description}
The last line of (\ref{complements}) is paired with the ghost
sector (\ref{k+1C}) which disappear from the cohomology. The
second line of (\ref{complements}) states that the traceless
component $\widehat{\,[{\cal R }]\,}$ of the derivatives of the
curvature tensors are in the cohomology since they span the
irreducible $\mathfrak{o}(n)$-modules labeled by
$[\begin{picture}(80,15)(-5,2)
\multiframe(0,6.5)(10.5,0){1}(60,7){}\put(65,6.5){k}
\multiframe(0,-1)(10.5,0){1}(30,7){}\put(50,-2){s}
\end{picture}]$ while the third line
of (\ref{complements}) together with (\ref{kphidoubletrace}) and (\ref{k-2F}) show that the derivatives of the Fronsdal
tensor depending on $\widetilde{\varphi}$ complete the generators
of the cohomology of $\gamma$.

\section{Proof of Theorem \ref{Killing}}\label{proof2}

\subsection{Unconstrained off-shell Killing tensor fields}\label{shproof}

If $\varepsilon_{\m_1\ldots\m_{s-1}}(x)$ is a formal power series
in $x\,$, then, by definition,
\begin{equation}\label{formpow}\varepsilon_{\m_1\ldots\,\m_{s-1}}(x)=\sum\limits_{t=0}^\infty
\l_{\m_1\ldots\,\m_{s-1}\,,\,\,\n_1\ldots\,\n_t}\,x^{\n_1}\ldots
\,x^{\n_t}\,, \end{equation} where the tensors
$\l_{\m_1\ldots\,\m_{s-1}\,,\,\,\n_1\ldots\,\n_t}$ are all
constant and each set of indices is symmetrized. If one acts with
$s-1$ partial derivatives on both sides of the off-shell
Killing-like equation (\ref{Killing}), then the resulting equation
is equivalent to\begin{equation}
\partial_{\n_1}\ldots\partial_{\n_s}\varepsilon_{\m_1\ldots\m_{s-1}}=
0 \label{Ks}\end{equation} because all terms in the decomposition
of the tensor product $\{\begin{picture}(60,15)(-5,2)
\put(25,11){s} \multiframe(0,2)(10.5,0){1}(8.5,7){$\cdot$}
\multiframe(9,2)(10.5,0){1}(30,7){$\cdots$}
\multiframe(39.5,2)(10.5,0){1}(8.5,7){$\cdot$}
\end{picture}\}\otimes\{\begin{picture}(50,15)(-5,2) \put(16,11){s-1}
\multiframe(0,2)(10.5,0){1}(40,7){}
\end{picture}\}
$ (where a dot in a box stands for a partial derivative) always
contain, as a subdiagram, the Young diagram
$\begin{picture}(60,15)(-5,2) \put(25,11){s}
\multiframe(0,2)(10.5,0){1}(40,7){}
\multiframe(40.5,2)(10.5,0){1}(8.5,7){$\cdot$}
\end{picture}$ corresponding
to the left-hand-side of (\ref{Killingequ}). In other words, the
left-hand-side of (\ref{Ks}) depends linearly on the
left-hand-side of (\ref{Killingequ}). Inserting (\ref{formpow}) in
(\ref{Ks}) leads to the fact that the tensors
$\l_{\m_1\ldots\,\m_{s-1}\,,\,\,\n_1\ldots\,\n_t}$ are zero for $t
\geq s$.
Then, substituting the resulting polynomial for
$\varepsilon_{\m_1\ldots\m_{s-1}}$ in the original Killing-like
equation (\ref{Killingequ}) gives the system of equations
\begin{equation}
\l_{(\m_1\ldots\,\m_{s-1}\,,\,\,\n_1)\n_2\ldots\,\n_t}=0\quad(t<s)
\label{systequ} \end{equation} which implies that the constant
tensor $\l_{\m_1\ldots\,\m_{s-1}\,,\,\,\n_1\ldots\,\n_t}$ belongs
to the $\mathfrak{gl}(n)$-module labeled by
$\{\begin{picture}(90,15)(0,2)
\multiframe(10,7.5)(10.5,0){1}(45,7){}\put(60,7.5){$s-1$}
\multiframe(10,0)(10.5,0){1}(25,7){}\put(41,-1){$t$}
\end{picture}\}$. This proves Lemma
\ref{lem}.\vspace{1mm}

Analogously, the point (i) of Theorem \ref{Killing} is a
straightforward consequence of the relations (\ref{rel}) applied
to any monomial of the schematic form $\textsc{M}^t
\textsc{P}^{s-1-t}$ which leads to a result identical to the
previous one in terms of irreducible $\mathfrak{gl}(n)$-modules.
The main point is that the $\mathfrak{gl}(n)$-irreducibility
conditions (\ref{rel}) are expressed in the ``antisymmetric"
convention for Young diagrams.
Alternatively, the isomorphism (i) follows from the
fact that the algebra of Killing tensors on any constant curvature space
is generated by the corresponding Killing vectors \cite{Killingconstcurvv}
(which are of course in one-to-one correspondence with the generators of the isometry algebra).

\subsection{Constrained on-shell Killing tensor fields}

The proof of the point (ii) of Theorem \ref{Killing} requires more
work and uses some general results on local Koszul-Tate cohomology
groups. Let us consider the on-shell (also called ``weak")
Killing-like equation (\ref{weaK}) for the constrained free
spin-$s$ gauge field theory:
\begin{equation}\partial_{(\m_1}\widehat{\varepsilon}_{\m_2\ldots\m_s)}\approx
0\,.\label{weaKc}\end{equation} As explained in the previous
subsection, if one acts with $s-1$ partial derivatives on both
sides of (\ref{weaKc}), then the resulting equation is equivalent
to\footnote{The equation (\ref{weakKs}) was conjectured for
arbitrary spin $s$ in \cite{Nazim}.}\begin{equation}
\partial_{\n_1}\ldots\partial_{\n_s}\widehat{\varepsilon}_{\m_1\ldots\m_{s-1}}\approx
0\,,\label{weakKs}\end{equation} because the partial derivative
preserves the stationary surface since the Fronsdal equation
(\ref{Fronsdalequ}) does not depend explicitly on $x$.

An application of the general theorems 6.2 and 6.4 of
\cite{Barnich:2000zw} to the constrained free spin-$s$ gauge field
theory of Fronsdal leads to the fact that the ``characteristic
cohomology group" in form degree zero is represented by the
constants (see for instance
\cite{Henneaux:1998rp,Boulanger:2000rq,mixed}). In other words,
$$\partial_\m f(x,[\widetilde{\varphi}])\approx
0\quad\Longleftrightarrow \quad f(x,[\widetilde{\varphi}])\approx
cst\,,
$$ where ``$cst$" stands for a constant independent of $x$ and the (partial
derivatives of the) gauge field $\widetilde{\varphi}$. Therefore,
(\ref{weakKs}) is equivalent to
$$\partial_{\n_1}\ldots\partial_{\n_{s-1}}\widehat{\varepsilon}_{\m_1\ldots\m_{s-1}}\approx
\l_{\m_1\ldots\,\m_{s-1}\,,\,\,\n_1\ldots\,\n_{s-1}}\,,$$ where
the tensor $\l_{\m_1\ldots\,\m_{s-1}\,,\,\,\n_1\ldots\,\n_{s-1}}$
is constant, each set of indices is symmetrized and the indices
$\mu$ are traceless. Consequently, the tensor
$$\widehat{\varepsilon}^{\,\prime}_{\m_1\ldots\m_{s-1}}:=\widehat{\varepsilon}_{\m_1\ldots\m_{s-1}}-
\l_{\m_1\ldots\,\m_{s-1}\,,\,\,\n_1\ldots\,\n_{s-1}}\,x^{\m_1}\ldots
x^{\m_{s-1}} \,,$$ satisfies
$$\partial_{\n_1}\ldots\partial_{\n_{s-1}}\widehat{\varepsilon}^{\,\prime}_{\m_1\ldots\m_{s-1}}\approx 0\,.$$
By repeating the previous argument $s-1$ times, we arrive at the
conclusion that
\begin{equation}\label{pow}\widehat{\varepsilon}_{\m_1\ldots\,\m_{s-1}}(x,[\widetilde{\varphi}])\approx
\sum\limits_{t=0}^{s-1}
\l_{\m_1\ldots\,\m_{s-1}\,,\,\,\n_1\ldots\,\n_t}\,x^{\n_1}\ldots
\,x^{\n_t}\,, \end{equation} where the tensors
$\l_{\m_1\ldots\,\m_{s-1}\,,\,\,\n_1\ldots\,\n_t}$ are constant,
symmetric in each set of indices and traceless in the indices
$\mu$. Substituting (\ref{pow}) in the (weak) Killing-like
equation (\ref{weaKc}), we obtain
\begin{equation}\sum\limits_{t=0}^{s-2}
(t+1)\,\l_{(\m_1\ldots\,\m_{s-1}\,,\,\,\m_s)\,\n_1\ldots\,\n_t}\,x^{\n_1}\ldots
\,x^{\n_t}\approx 0\,. \label{equw}\end{equation} Taking into
account the fact that the left-hand-side of the (weak) equality
(\ref{equw}) depends only on the space-time variable $x$ while the
Fronsdal equation (\ref{Fronsdalequ}) is linear in the jet space
variables $[\widetilde{\varphi}]$, we arrive at the conclusion
that the constant tensors $\l$ satisfy the (strong) conditions
(\ref{systequ}).

To end up the proof of the last point of Theorem
\ref{Killing}, we use the general property that tensors
irreducible under $\mathfrak{gl}(n)$ and traceless in the indices
corresponding to the largest row are traceless in all their
indices (see also Lemma 4.5 of \cite{BGST} for another proof of the symmetry
properties of traceless Killing tensors).



\begin{thebibliography}{99}

\bibitem{Fronsdal:1978rb}
C.~Fronsdal,
Phys.\ Rev.\ D {\bf 18} (1978) 3624.

\bibitem{BCIV}X. Bekaert, S. Cnockaert, C. Iazeolla and M.A.
Vasiliev, ``Nonlinear Higher Spin Theories in Various
Dimensions'', in the proceedings of the First Solvay Workshop on
Higher-Spin Gauge Theories (Brussels, Belgium, May 2004) [{\tt
hep-th/0503128}].

\bibitem{deWit:1979pe}
B.~de Wit and D.~Z.~Freedman,
Phys.\ Rev.\ D {\bf 21} (1980) 358.

\bibitem{Vframe} M.~A.~Vasiliev, Sov.\ J.\ Nucl.\ Phys.\ {\bf 32} (1980)
439 [Yad.\ Fiz.\  {\bf 32} (1980) 855]\,; Fortsch.\ Phys.\  {\bf
35} (1987) 741\,; \\
V.~E.~Lopatin and M.~A.~Vasiliev, Mod.\ Phys.\ Lett.\ A {\bf 3}
(1988) 257.

\bibitem{Francia:2002aa}
D.~Francia and A.~Sagnotti,
Phys.\ Lett.\ B {\bf 543} (2002) 303 [{\tt hep-th/0207002}].

\bibitem{Bouatta:2004kk}
D.~Sorokin, ``Introduction to the classical theory of higher
spins'', in the proceedings of XIX Max Born Symposium (Wroclaw,
Poland, October 2004) pp. 172-202
[{\tt hep-th/0405069}];\\
N.~Bouatta, G.~Compere and A.~Sagnotti, ``An introduction to free
higher-spin fields'', in the proceedings of the First Solvay
Workshop on Higher-Spin Gauge Theories (Brussels, Belgium, May
2004) [{\tt hep-th/0409068}].

\bibitem{DD}
T.~Damour and S.~Deser,
Annales Poincar\'e Phys.\ Th\'eor.\  {\bf 47} (1987) 277.

\bibitem{DuboisV}
P.J.~Olver, ``Differential hyperforms {I}", Univ. of Minnesota
report 82-101\,;\\
M.~Dubois-Violette and M.~Henneaux, Lett.\ Math.\ Phys.\ {\bf 49}
(1999) 245 [{\tt math.qa/9907135}];
Commun.\ Math.\ Phys.\  {\bf 226} (2002) 393 [{\tt math.qa/0110088}].

\bibitem{BB2}
X.~Bekaert and N.~Boulanger,
Phys.\ Lett.\ B {\bf 561} (2003) 183 [{\tt hep-th/0301243}];
``Mixed symmetry gauge fields in a flat background'', in the
proceedings of the International Seminar on Supersymmetries and
Quantum Symmetries ``SQS 03" (Dubna, Russia, July 2003) [{\tt
hep-th/0310209}].

\bibitem{Henneauxbook}
M.~Henneaux and C.~Teitelboim, \textit{Quantization of Gauge
Systems} (Princeton University Press, 1992).

\bibitem{BBetal}
X. Bekaert, N. Boulanger and S. Cnockaert, in preparation.

\bibitem{BB1}
X.~Bekaert and N.~Boulanger,
Commun.\ Math.\ Phys.\  {\bf 245} (2004) 27 [{\tt
hep-th/0208058}];
Class.\ Quantum\ Grav.\  {\bf 20} (2003) S417 [{\tt
hep-th/0212131}].

\bibitem{Shaynk}
O.~V.~Shaynkman and M.~A.~Vasiliev,
Theor.\ Math.\ Phys.\  {\bf 123} (2000) 683 [Teor.\ Mat.\ Fiz.\
{\bf 123} (2000) 323] [{\tt hep-th/0003123}];
Theor.\ Math.\ Phys.\  {\bf 128} (2001) 1155 [Teor.\ Mat.\ Fiz.\
{\bf 128} (2001) 378] [{\tt hep-th/0103208}];\\
O.~A.~Gelfond and M.~A.~Vasiliev,
[{\tt hep-th/0304020}].


\bibitem{Eisen}
L.~P.~Eisenhart, \emph{Riemannian Geometry} (Princeton University
Press, 1966) p. 128.

\bibitem{Killingconstcurvv}
G. Thompson,
J. Math. Phys. {\bf 27} (1986) 2693.

\bibitem{Killingtensorsolution}
T. Wolf,
Comp.\ Phys.\ Comm.\ {\bf 115}  (1998) 316;\\
R.~G.~McLenaghan, R.~Milson and R.~G.~Smirnov, C.\ R.\ Acad.\ Sci.\ Paris, Ser. {\bf I 339} (2004) 621.

\bibitem{Barnich:2000zw}
G.~Barnich, F.~Brandt and M.~Henneaux,
Phys.\ Rept.\  {\bf 338} (2000) 439 [{\tt hep-th/0002245}].

\bibitem{Eastwood}
M.~G.~Eastwood, ``Higher symmetries of the Laplacian''  
[{\tt hep-th/0206233}].

\bibitem{Valgebra}
M.~A.~Vasiliev, Phys.\ Lett.\ B {\bf 567} (2003) 139 [{\tt
hep-th/0304049}].

\bibitem{Berends}
F.~A.~Berends, G.~J.~H.~Burgers and H.~van Dam,
Nucl.\ Phys.\ B {\bf 271} (1986) 429.

\bibitem{cm} S.~R.~Coleman and J.~Mandula,
Phys.\ Rev.\  {\bf 159} (1967) 1251;\\
R.~Haag, J.~T.~Lopuszanski and M.~Sohnius, Nucl.\ Phys.\ B {\bf
88}, 257 (1975).

\bibitem{B}
X. Bekaert, in preparation.

\bibitem{Henneaux:1989}
M.~Henneaux,
Nucl. Phys. Proc. Suppl. {\bf 18A} (1990) 47.

\bibitem{BGST}
G.~Barnich, M.~Grigoriev, A.~Semikhatov and I.~Tipunin, ``Parent
field theory and unfolding in BRST first-quantized terms'' [{\tt hep-th/0406192}].

\bibitem{Mik}
A. Mikhailov, ``Notes On Higher Spin Symmetries'' [{\tt hep-th/0201019}].

\bibitem{Sagnotti:2005}
A.~Sagnotti, E.~Sezgin and P.~Sundell, ``On higher spins with a
strong $Sp(2,{\mathbb R})$ condition'', in the proceedings of the
First Solvay Workshop on Higher-Spin Gauge Theories (Brussels,
Belgium, May 2004) [{\tt hep-th/0501156}].

\bibitem{Vasiliev:1999}
D.~Anselmi,
Nucl.\ Phys.\ B {\bf 541} (1999) 323
[{\tt hep-th/9808004}];
Class.\ Quantum\ Grav.\  {\bf 17} (2000) 1383
[{\tt hep-th/9906167}];\\
M.~A.~Vasiliev, ``Higher spin gauge theories: Star-product and AdS
space'', [{\tt hep-th/9910096}] in M. Shifman ed., \textit{The many faces of the
superworld} (World Scientific, 2000)
Section 2.

\bibitem{Vunf}
M.~A.~Vasiliev, ``Actions, charges and off-shell fields in the
unfolded dynamics approach'' [{\tt hep-th/0504090}].

\bibitem{Nazim}
N. Bouatta, ``Approche BRST des syst\`emes invariants de jauge \`a
spins \'elev\'es'', Master Degree Thesis defended at the Free
University of Brussels (Academic year 2003-2004).

\bibitem{BaBo}
G. Barnich, private communication.

\bibitem{Henneaux:1998rp}
M.~Henneaux and B.~Knaepen,
Nucl.\ Phys.\ B {\bf 548} (1999) 491
[{\tt hep-th/9812140}].

\bibitem{Boulanger:2000rq}
N.~Boulanger, T.~Damour, L.~Gualtieri and M.~Henneaux,
Nucl.\ Phys.\ B {\bf 597} (2001) 127 [{\tt hep-th/0007220}].

\bibitem{mixed}
J.~A.~Garcia and B.~Knaepen,
Phys.\ Lett.\ B {\bf 441} (1998) 198
[{\tt hep-th/9807016}];\\
X.~Bekaert, N.~Boulanger and M.~Henneaux,
Phys.\ Rev.\ D {\bf 67} (2003) 044010
[{\tt hep-th/0210278}];\\
N.~Boulanger and S.~Cnockaert,
JHEP {\bf 0403} (2004) 031 [{\tt hep-th/0402180}];\\
X.~Bekaert, N.~Boulanger and S.~Cnockaert,
J. Math. Phys. {\bf 46} (2005)
012303 [{\tt hep-th/0407102}].

\bibitem{Littlewood}
D.E. Littlewood, {\it The theory of group characters} (Clarendon,
1940); 
G.~R.~E.~Black, R.~C.~King and B.~G.~Wybourne, J.\ Phys.\ A:\ Math.\ Gen.\ {\bf 16} (1983) 1555; 
S.~A.~Fulling, R.~C.~King, B.~G.~Wybourne and C.~J.~Cummins,
Class.\ Quantum\ Grav.\  {\bf 9} (1992) 1151.

\bibitem{King}
R.C. King, J. Phys. A: Math. Gen. {\bf 8} (1975) 429.

\end{thebibliography}
\end{document}